\documentclass[9pt]{elife}

\usepackage{lipsum} 
\usepackage[version=4]{mhchem}
\usepackage{siunitx}
\DeclareSIUnit\Molar{M}
\usepackage{rotating} 

\usepackage{todonotes}
\usepackage{caption}
\usepackage{subcaption}
\usepackage{graphicx}

\newcounter{boxfig}
\newcommand{\BOXFIG}[2][]{%
  \ifstrequal{#1}{}{}{\autoref{box:#1}--}\autoref{boxfig:#2}%
}

\title{Computational reproducibility of Jupyter notebooks from biomedical publications}

\author[1,2 *\authfn{1}]{Sheeba Samuel}
\author[3,4,5*\authfn{1}]{Daniel Mietchen}
\affil[1]{Heinz-Nixdorf Chair for Distributed Information Systems,
Friedrich Schiller University Jena, Germany}
\affil[2]{Michael Stifel Center Jena, Germany}
\affil[3]{Ronin Institute, Montclair, New Jersey, United States}
\affil[4]{Institute for Globally Distributed Open Research and Education (IGDORE)}
\affil[5]{FIZ Karlsruhe~---~Leibniz Institute for Information Infrastructure, Berlin, Germany}

\corr{sheeba.samuel@uni-jena.de}{SS}
\corr{daniel.mietchen@ronininstitute.org}{DM}

\contrib[\authfn{1}]{These authors contributed equally to this work}

\begin{document}

\maketitle

\begin{abstract}
Jupyter notebooks allow to bundle executable code with its documentation and output in one interactive environment, and they represent a popular mechanism to document and share computational workflows, including for research publications. 
Here, we analyze the computational reproducibility of 9625 Jupyter notebooks from 1117 GitHub repositories associated with 1419 publications indexed in the biomedical literature repository PubMed Central. 
8160 of these were written in Python, including 4169 that had their dependencies declared in standard requirement files and that we attempted to re-run automatically. 
For 2684 of these, all declared dependencies could be installed successfully, and we re-ran them to assess reproducibility. Of these, 396 notebooks ran through without any errors, 
including 245 that produced results identical to those reported in the original. 
Running the other notebooks resulted in exceptions.
We zoom in on common problems and practices, highlight trends and discuss potential improvements to 
Jupyter-related workflows associated with biomedical publications.
\end{abstract}

\newpage
\tableofcontents
\listoffigures
\listoftables
\newpage

\section{Introduction} 
\label{sec:Intro}

Many factors contribute to the progress of scientific research, including the precision, scale, and speed at which research can be performed and shared and the degree to which research processes and their outcomes can be trusted \citep{siebert2015point,contera2021communication}. This trust, in turn, and the credibility that comes with it, are a social construct that depends on past experience or proxies to it \citep{gray2012understanding,kroeger2018scientific,jamieson2019Signaling}.
A good proxy here is reproducibility, at least in principle \citep{hsieh2018enhancing}: if a study addressing a particular research question can be re-analyzed and that analysis leads to the same conclusions as the original study, then these conclusions can generally be more trusted than if the conclusions differ between the original and the replication study.

\subsection{Reproducibility issues in contemporary research}
\label{sec:Reproducibility-issues}

Over recent years, the practical replicability of published research has come into focus and turned into a research area in and of itself 
\citep{peng2015thereproducibility, samuel2021understanding}.
As a result, systematic issues with reproducibility have been the subject of many publications in various research fields as well as 
prominent mentions 
in the mass media \citep{theeconomist2013trouble}.
These research fields range from psychology \citep{simmons2011false} to cell culture \citep{hussain2013reproducible,bairoch2018cellosaurus} to ecology \citep{kelly2019rate}, geosciences \citep{ledermann2021towards} and beyond and include software-affine domains such as health informatics \citep{coiera2018does}, 
human-computer interactions \citep{hinsen2018verifiability}, 
artificial intelligence \citep{hutson2018artificial}, software engineering \citep{shepperd2018role} and
research software \citep{crick2017reproducibility}. 
This is often framed in terms of a ``reproducibility crisis'' \citep{baker20161500}, though that may not necessarily be the most productive approach to addressing the underlying issues \citep{hunter2017the,fanelli2018opinion,guttinger2020limits}. In more practical terms,  \citet{napflin2019genomics} observe that ``appropriate workflow documentation is essential''. 

\subsection{Terminology}
\label{sec:Terminology}
Within this broader context, distinctions between replicability, reproducibility, and repeatability are often important or even necessary \citep{meng2020reproducibility} but not consistently made in the literature \citep{plesser2017reproducibility}.
A potential solution to this confusion is the proposed distinction  \citep{goodman2016what} between 
\emph{Methods reproducibility} (providing enough detail about the original study that the procedures and data can be repeated exactly), 
\emph{Results reproducibility} (obtaining the same results when matching the original procedures and data as closely as possible) and
\emph{Inferential reproducibility} (leading to the same scientific conclusions as the original study, either by reanalysis or by independent replication).

In the following, we will concentrate on ``Methods reproducibility in computational research'', i.e.\ using the same code on the same data source. For this, we will use the shorthand ``Computational reproducibility''. In doing so, we are conscious that the ``same code'' can yield different results depending on the execution environment and that the ``same data source'' might actually mean different data if the data source is dynamic or if the code involves manipulating the data in a way that changes over time. 
We are also aware that the shorthand ``Computational reproducibility'' can also be applied, e.g.,
to ``Results reproducibility in computational research'' in cases where the algorithm described for the original study was re-implemented in a follow-up study. For instance, \citet{burlingame2021toward} were striving for \emph{Results reproducibility} when they re-implemented the PhenoGraph algorithm~-- which originally only ran on CPUs~-- such that it could be run on GPUs and thus at higher speed.
However, \emph{Results reproducibility} is not the focus of our study.

\subsection{Computational reproducibility in biomedical research}
\label{sec:Computational-Reproducibility}
In light of the reproducibility issues outlined above, there have been calls for better standardization of biomedical research software~-- see \citet{russell2018large} for an example.
In line with such standardization calls, a number of guidelines or principles
to achieve
methods reproducibility in several computational research contexts have been proposed. For instance, \cite{sandve2013ten}, \cite{gil2016toward} and \cite{Willcox2021ReSearchOps} laid out principles for reproducible computational research in general.
In a similar vein, \cite{gruning2018practical} and \cite{brito2020recommendations} looked at specifics of computational reproducibility in the life sciences, \cite{nust2020ten} explored the use of Docker~-- a containerization tool~-- in reproducibility contexts, 
and \cite{trisovic2022large} looked at the reproducibility of R scripts archived in an institutional repository, while
\cite{rule2019ten}, \cite{pimentel2019a} as well as \cite{wang2020restoring,willis2020developing} and \cite{wang2020better} zoomed in on Jupyter notebooks, a popular file format for documenting and sharing code.
While most of these are language agnostic, language-specific approaches to computational reproducibility have also been outlined, e.g.\  for Python
 \citep{halchenko2021datalad}.
 
 However, compliance with such standards and guidelines is not a given \citep{russell2018large,rule2018exploration,pimentel2021understanding}, so we set out to measure it specifically for Jupyter notebooks in the life sciences
 and to explore options to bridge the gap between recommended and actual practice.
 In order to do so, we mined a popular repository of biomedical fulltexts (PubMed Central) for mentions of
 Jupyter notebooks alongside mentions of a popular 
 repository for open-source software (GitHub). 
 
\subsection{PubMed Central} 
\label{sec:PMC}
PubMed Central (PMC)\footnote{\url{https://www.ncbi.nlm.nih.gov/pmc/}} is a literature repository containing full texts of biomedical articles.
At the time of writing, it contained about 7.5 million articles. 
Founded in the context of the Open Access mandate issued by the National Institutes of Health (NIH) in the United States \citep{roberts2001pubmed},
PMC is operated by the National Center for Biotechnology Information (NCBI), a branch of the National Library of Medicine (NLM), which is part of the NIH.
PMC hosts the articles using the Journal Article Tagging Suite (JATS), an XML standard, and makes them available for manual and programmatic access in various ways, of which we used the Entrez API \citep{sayers2010ageneral}.

\subsection{GitHub} 
\label{sec:GitHub}
GitHub\footnote{\url{https://github.com/}} is a website that combines git-based version control with support for collaboration and automation. It is a popular place for sharing software and developing it collaboratively, including for Jupyter notebooks \citep{rule2018exploration} and for code associated with research articles available through PubMed Central \citep{russell2018large}.

\subsection{Jupyter}
\label{sec:Jupyter}
Jupyter notebooks\footnote{\url{https://jupyter.org/}} \citep{kluyver2016jupyter,granger2021jupyter} are a computing environment in which code, code documentation, and output of the code can be explored interactively. They have become a popular mechanism to share computational workflows in a variety of fields \citep{kluyver2016jupyter}, including astronomy
\citep{randles2017using,wofford2019jupyter} and biosciences \citep{schroder2019reproducible}. Here, we build on past studies of the reproducibility of Jupyter notebooks \citep{rule2018exploration, pimentel2019a} and 
analyze Jupyter notebooks available through GitHub repositories associated with publications available through the biomedical literature repository PubMed Central.

\subsection{Jupyter and reproducibility}
\label{sec:Jupyter-reproducibility}
Jupyter notebooks can, in principle, be used to enhance reproducibility, and they are often presented as such, yet using them does not automatically confer reproducibility to the code they contain.
Several studies have been conducted in recent years to explore the reproducibility of Jupyter Notebooks.
A recent one has investigated the reproducibility of Jupyter notebooks associated with five publications from the PubMed Central database \citep{schroder2019reproducible}. In their reproducibility analysis, they looked for the presence of notebooks, source code artifacts, documentation of the software requirements, and whether the notebooks can be re-executed with the same results. According to their results, the authors successfully reproduced only three of 22 notebooks from five publications.
Rule et al. \citep{rule2018exploration} explored 1 million notebooks available on GitHub. In their study, they explored repositories, language, packages, notebook length, and execution order, focusing on on the structure and formatting of computational notebooks. As a result, they provided ten best practices to follow when writing and sharing computational analyses in Jupyter Notebooks \citep{rule2019ten}.
Another study \citep{pimentel2021understanding} focused on the reproducibility of 1.4 million notebooks collected from GitHub. It provides an extensive analysis of the factors that impact reproducibility based on Jupyter notebooks.
Chattopadhyay et al. \citep{chattopadhyay2020s} reported on the results of a survey conducted among 156 data scientists on the difficulties when working with notebooks.
Other studies focus on best practices on writing and sharing Jupyter notebooks \citep{rule2019ten, pimentel2021understanding, willis2020developing, wang2020better}.
As a result, tools have been developed to support provenance and reproducibility in Jupyter Notebooks \citep{chirigati2013reprozip, boettiger2015Docker, samuel2018provbook, jupyter2018binder}. 
Cases where Jupyter notebooks have played a key role in some actual replication attempts have also begun to appear in the literature. For instance, \citet{baker2019quantification} assembled a Jupyter notebook as part of a published correction.
Shortly after we had created our corpus, a paper was published with a Jupyter notebook that enabled  others to reproduce the computational workflows, ultimately leading to the retraction of the original work, as detailed in \citet{meyerowitz2021impact}.

\subsection{Wikidata}
\label{sec:Wikidata}
Wikidata is a cross-disciplinary and multilingual database through which a global community curates FAIR and open data to serve as general reference information \citep{waagmeester2020wikidata,rutz2022LOTUS}. This includes information about key elements of the research ecosystem, from researchers to research fields and research organizations, from methods to datasets, software and publications. 
This information can then be explored in various ways, e.g. through the visualization tool Scholia \citep{nielsen2017scholia}, which provides profiles for different types of entities or relationships.
For entities of the type Jupyter notebook (known to Wikidata as \href{https://www.wikidata.org/wiki/Q70357595}{Q70357595}), the most relevant profile types are those for a topic\footnote{\href{https://scholia.toolforge.org/topic/Q70357595}{https://scholia.toolforge.org/{\textbf{topic}/Q70357595}}}, a software \footnote{\href{https://scholia.toolforge.org/software/Q70357595}{https://scholia.toolforge.org/\textbf{software}/Q70357595}} or a resource used\footnote{ \href{https://scholia.toolforge.org/use/Q70357595}{https://scholia.toolforge.org/\textbf{use}/Q70357595}}.

\subsection{Environmental footprint} 
\label{sec:Environmental-footprint}
Computations ultimately require physical resources, and both the production and the use of these resources can have a considerable environmental footprint \citep{lannelongue2021green}. The more reproducible some workflows become, the more accurate their environmental footprint can be assessed \citep{taddeo2021artificial}. 
This can then lead to an optimization of the environmental footprint, especially since it often correlates with the financial footprint of using computational resources \citep{schwartz2020green}.
One of our aims in this study is thus to get an overview of the contribution of Jupyter-based workflows to the environmental footprint of biomedical research involving computation. This is in line with the recommendation in \citet{lannelongue2021ten} to integrate routine environmental footprint assessment into research practice.

\section{Methods}
\label{sec:Methods}

\subsection{Pipeline} 
\label{sec:Pipeline}

\begin{figure}[ht!]
\centering
\includegraphics[width=1.0\textwidth]{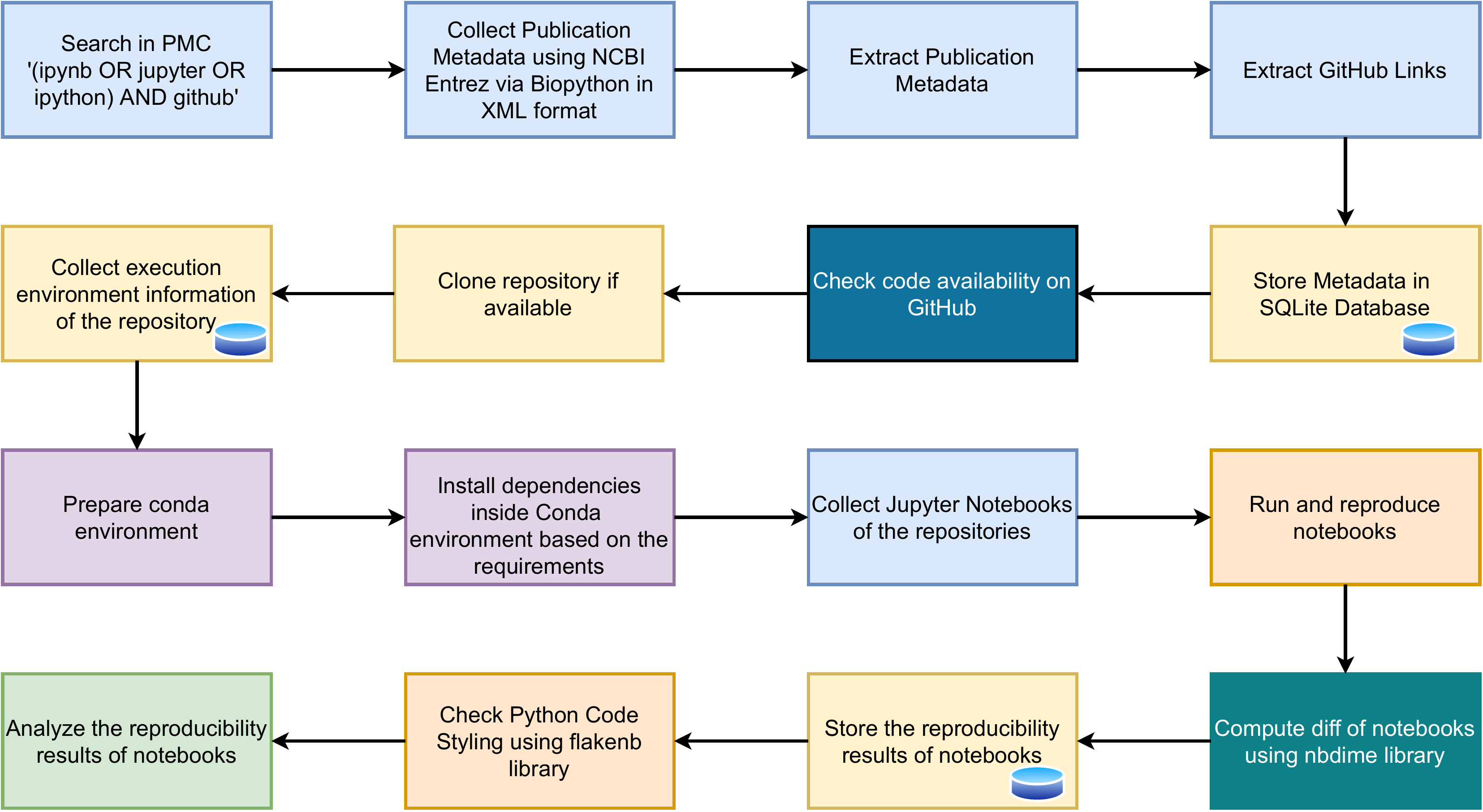}
\caption[Workflow]{Fully automated workflow used for assessing the reproducibility of Jupyter notebooks from publications indexed in PubMed Central: the PMC search query resulted in a list of article identifiers that were then used to retrieve the full-text XML, from which publication metadata and GitHub links were extracted and entered into an SQLite database. If the links pointed to valid GitHub repositories containing valid Jupyter notebooks, then metadata about these were gathered, and the Python-based notebooks were run with all identifiable dependencies, and their results analyzed with respect to the originally reported ones.
}
\label{fig:Figure_MethodsWorkflow}
\end{figure}

In this section, we describe the key steps of the pipeline we used for assessing the reproducibility of Jupyter notebooks associated with publications extracted from PubMed Central. 
The driver file for running the workflow is documented in r0\_main.py and the driver notebook for the analysis of the collected data is documented in the notebook named ``Index.ipynb''.
Figure \ref{fig:Figure_MethodsWorkflow} provides an overview of the workflow used in this study.

We used the \textit{esearch} function to search PMC for Jupyter notebooks on 24\textsuperscript{th} February, 2021.
We looked for publications that mentioned GitHub together with either the string ``Jupyter'' or some closely associated ones, namely ``ipynb'' (the file ending/extension of Jupyter notebooks) or ``iPython'' (the name of a precursor to Jupyter).
The search query used was ``(ipynb OR jupyter OR ipython) AND github''.
Based on the primary PMC IDs received from the \textit{esearch} utility, we retrieved records in the XML format using the \textit{efetch} function and collected the publication metadata from PMC \citep{roberts2001pubmed} using NCBI Entrez utilities via Biopython \citep{cock2009biopython}.

In the next step, we processed the XML fetched from PMC.
We used an SQLite database\footnote{\url{https://www.sqlite.org}} for storing all the data related to our pipeline.
We collected information on journals and articles.
We first extracted information about the journal.
For this, we created a database table 
for the journal and extracted the ISSN\footnote{\url{https://www.issn.org/}} (
International Identifier for serials), the journal title, the NLM's (National Library of Medicine) abbreviated journal title, and the ISO\footnote{\url{https://www.iso.org}} (International Organization for Standardization) abbreviation.

We then created a database table for the articles and populated it with article metadata.
The metadata includes the article name, Pubmed ID, PMC ID, Publisher id and name, DOI, subject, the dates when the article was received, accepted, and published, the license, the copyright statement, keywords, and the GitHub repositories mentioned in the publication.
For each article, we also extracted the Medical Subject Headings (MeSH terms)\footnote{\url{https://www.ncbi.nlm.nih.gov/mesh}} to get the subject area of the article.

To extract the GitHub repositories mentioned in each article, we looked for mentions of GitHub links anywhere in the article, 
including the abstract, the article body, data availability statement, and supplementary information.
GitHub links were available in different formats.
We normalized them to the standard format 'https://github.com/\{username\}/\{repositoryname\}'.
For example, we extracted the GitHub repository from nbviewer\footnote{\url{https://nbviewer.org/}} links and transformed its
representation to the standard format.
We excluded 692 GitHub links that mentioned only the username or organization name or github pages and not a specific repository name.
After preprocessing and extracting GitHub links from each article, we added the GitHub repositories to the database table for the corresponding articles.
Likewise, we linked the article's entry in the table to the journal where it was published.
We also collected information on the authors of the article in a separate database table: we created an author database table, extracted the first and last name, ORCID, email, and connected these data to the corresponding entries in the article table. 

Based on the GitHub repository name collected from the article, we checked whether these repositories were available at the original link or not. 
If the repository existed, we cloned it (ignoring branches, i.e.\ just taking the base one, which is usually called ``main'') and collected information about the repositories using the GitHub REST API\footnote{\url{https://docs.github.com/en/rest/guides/getting-started-with-the-rest-api}}.
On that basis, we created a repository database table.
For each GitHub repository, an entry is created in the table and connected to the article where it is mentioned.
We collected the execution environment information by looking into the dependency information declared in the repositories in terms of files like \textit{requirements.txt}, \textit{setup.py} and \textit{pipfile}.
Additional information for each repository is also collected from the GitHub API. 
This includes the dates of the creation, updates, or pushes to the repository, and the programming languages used in each repository. 
Further information includes the number of subscribers, forks, issues, downloads, license name and type, total releases, and total commits after the respective dates for when the article was published, accepted, and received.
After collecting and creating these data tables, we ran a pipeline to collect the Jupyter notebooks contained in the GitHub repositories. 
The code for the pipeline is adapted from \citep{pimentel2019a, samuel2021reproducemegit}.
Hence, the method to reproduce the notebooks in this study is similar to \citep{pimentel2019a}.
For each notebook, we collected information on the name, nbformat, kernel, language, number of different types of cells, and the maximum execution count number.
We extracted the source and output of each cell for  further analysis.
Using Python Abstract Syntax Tree (AST)\footnote{\url{https://docs.python.org/3/library/ast.html}} the pipeline extracted information on the use of modules, functions, classes, and imports.

After collecting all the required information for the execution of Python notebooks from the repositories, we prepared a Conda\footnote{\url{https://docs.conda.io/en/latest/}} environment based on the python version declared in the notebook.
Conda is an open source package and environment management system which helps users to easily find and install packages and create, save, load and switch between environments.
The pipeline then installed all the dependencies collected from the corresponding files like \textit{requirements.txt}, \textit{setup.py} and \textit{pipfile} inside the Conda environment. 
For the repositories that did not provide any dependencies using the above mentioned files, the pipeline executed the notebooks by installing all the anaconda dependencies\footnote{\url{https://docs.anaconda.com/anaconda/packages/pkg-docs/}}.
Anaconda is a Python and R distribution which provides data science packages including \textit{scikit-learn}, \textit{numpy}, \textit{matplotlib}, and \textit{pandas}.

We used the nbdime\footnote{\url{https://github.com/jupyter/nbdime}} 
library from Project Jupyter to compute diffs of the notebooks.
We used the tools adapted from \citep{pimentel2019a, samuel2021reproducemegit}.
The code from \citep{pimentel2019a} provides a basis for reproducing Jupyter notebooks from GitHub repositories.
The ReproduceMeGit \citep{samuel2021reproducemegit} extended from \citep{pimentel2019a}, is a visualization tool for analyzing the reproducibility of Jupyter Notebooks, along with provenance information of the execution.
ReproduceMeGit provides the difference between the results of the executions of notebooks using the nbdime library.
These two tools provide the basis for our code for the reproducibility study.

After collecting the notebooks, we also ran a Python code styling check using the flakenb\footnote{\url{https://github.com/s-weigand/flake8-nb}} library on the notebooks, since code styling consistency is a potential indicator for the extent of care that went into a given piece of software.
The flakenb library is a tool for code style guide enforcement for notebooks.
It helps to check code against some of the style conventions in PEP 8\footnote{\url{https://www.python.org/dev/peps/pep-0008/}}, a style guide for Python code.
The flakenb library provides an \textit{ignore} flag to ignore some specified errors.
In this study, we did not use this flag and collected all errors detected by the library.
For the styling of notebooks, we collected information on the pycode styling error code and description\footnote{\url{https://pycodestyle.pycqa.org/en/latest/intro.html}}.

\subsection{Reproduction}
\label{sec:Reproduction}

The complete pipeline was run on the Friedrich Schiller University Ara Cluster\footnote{\url{https://wiki.uni-jena.de/pages/viewpage.action?pageId=22453005}}.
The computational experiments were performed on a Skylake Standard Node (2x Intel Xeon Gold 6140 18 Core 2,3 GHz, 192 GB RAM).
This node has two CPUs, each with 18 cores, and 192 GB RAM in total.
The complete pipeline ran in 117 hours and 52 minutes from 24\textsuperscript{th}-28\textsuperscript{th} February 2021.
We then used the website \url{https://green-algorithms.org} v2.2 \cite{lannelongue2021green} to estimate that the pipeline run drew 47.38 kWh. Based in Germany, this has a carbon footprint of 16.05 kg CO2e, which is equivalent to 17.51 tree-months.

\section{Results} 
\label{sec:Results}

In this section, we present the results of our study on analyzing the computational reproducibility of Jupyter notebooks from biomedical publications.
We extracted metadata from 1419 publications from PubMed Central. 
These articles had been published in 373 journals and had 2398 mentions of GitHub repository links.
At the time of data collection, 49 GitHub repositories mentioned in the articles were not accessible, returning a ``page not found'' error instead. 
Out of 2177 unique and valid GitHub repositories cloned, only 1117 had one or more Jupyter notebooks.
From these repositories, a total of 9625 Jupyter notebooks were downloaded for further reproducibility analysis.\\
\subsection{General statistics of our study}
\label{sec:GeneralStatistics}

\begin{figure}[!htb]
\includegraphics[width=0.80\linewidth]{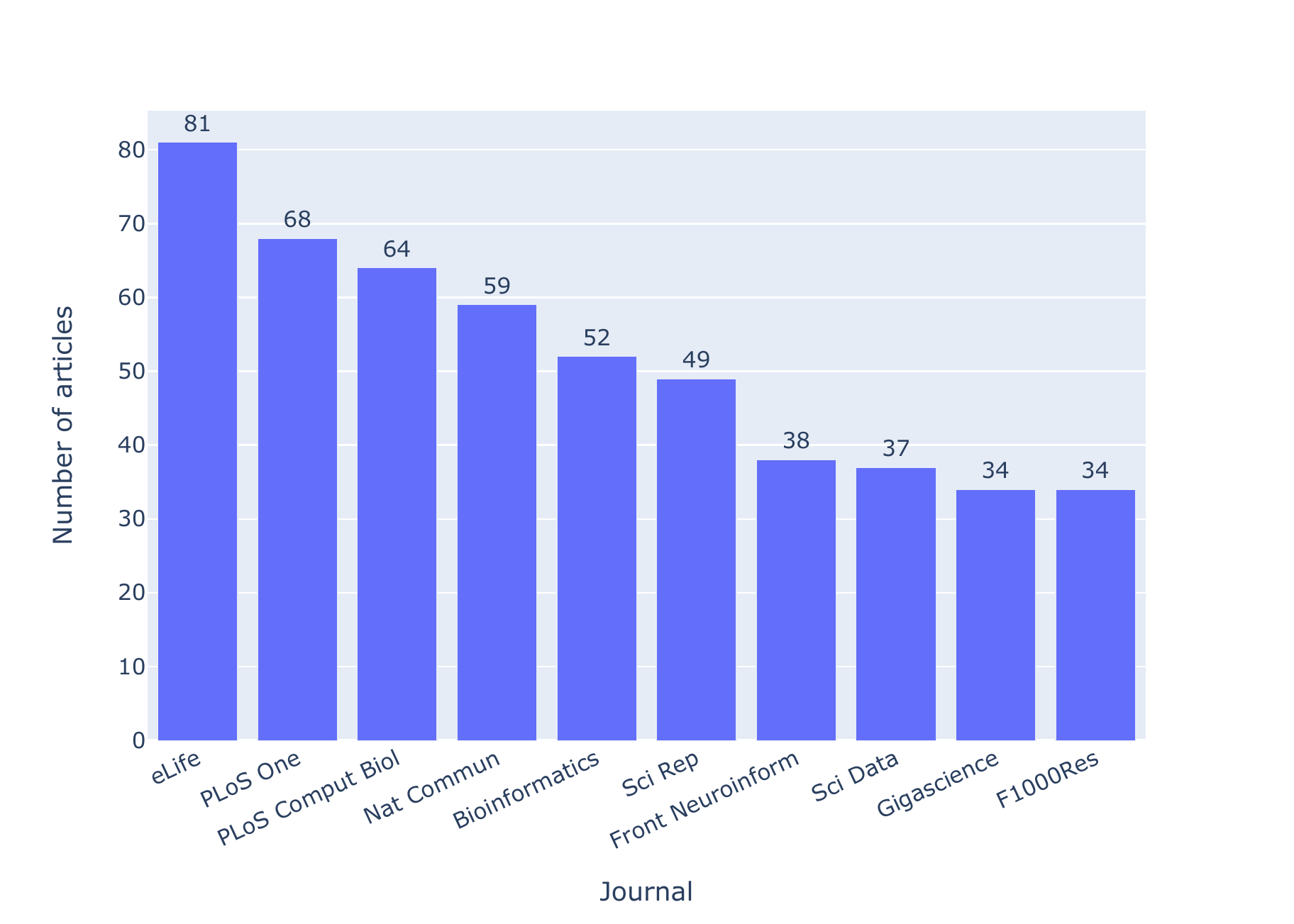}
\caption[Journals with the highest number of articles that had a valid GitHub repository and at least one Jupyter notebook.]{Journals with the highest number of articles that had a valid GitHub repository and at least one Jupyter notebook. In the figures, journal names are styled as in the XML files we parsed, e.g. (``PLoS Comput Biol''). In the text, we use the full name in its current styling, e.g.  ``PLOS Computational Biology''. 
}
\label{fig:Figure_top_journals_with_articles}
\end{figure}

\begin{figure}[!htb]
\includegraphics[width=\hsize]{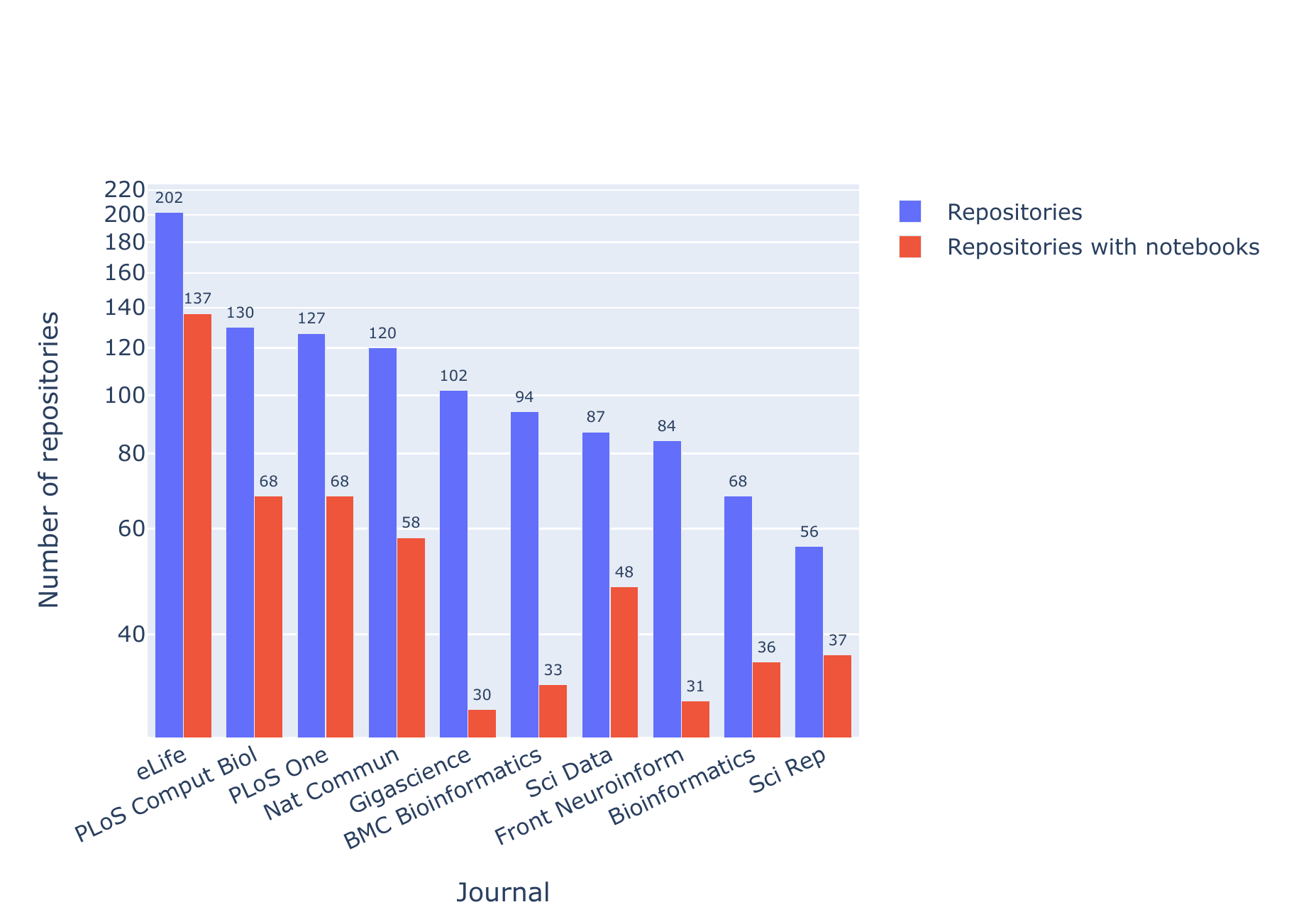}
\caption{Journals by the number of GitHub repositories and by the number of GitHub repositories with at least one Jupyter notebook.}
\label{fig:Figure_top_journals_repositories_with_without_notebooks}
\end{figure}
\FIG{Figure_top_journals_with_articles} shows the top ten journals with the highest number of articles that had a valid GitHub repository with at least one Jupyter notebook. 
\FIG{Figure_top_journals_repositories_with_without_notebooks} shows the journals by the number of GitHub repositories and repositories with Jupyter notebooks. 
The journal \textit{eLife} 
topped the list in both the rankings, 
which is why we chose to submit our manuscript there.
It was followed by \textit{PLOS ONE} and \textit{PLOS Computational Biology}.
The ratio of notebooks per GitHub repository varies across journals, with the range being between 3.4:1 in \textit{GigaScience}, 2:1 in \textit{Nature Communications},
and 1.5:1 in \textit{Scientific Reports}.
From the 1117 repositories with Jupyter notebooks, 290 (25.9\%) of repositories had one Jupyter notebook, 462 (41.4\%) had two notebooks, and 249 (22.3\%) had ten or more notebooks. 
6,782 (70.4\%) of the notebooks belonged to repositories with ten or more notebooks.

\begin{figure}[!htb]
\includegraphics[width=0.8\linewidth]{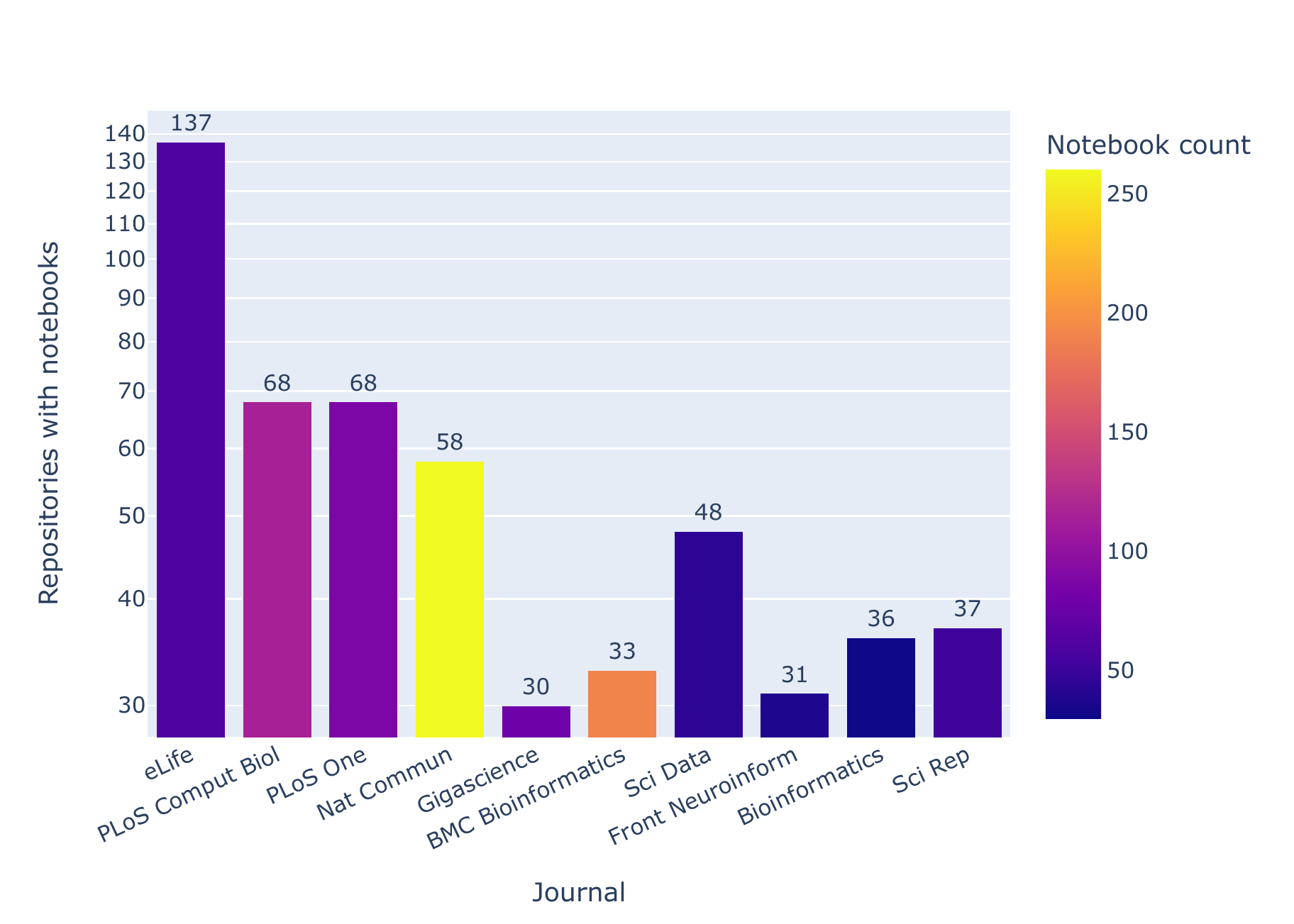}
\caption[Journals by number of GitHub repositories with Jupyter notebooks.]{Journals by number of GitHub repositories with Jupyter notebooks. For each journal, the notebook count gives the maximum number of notebooks within a repository associated with an article published in the journal.}
\label{fig:Figure_top_journals_with_repositories_notebooks}
\end{figure}

\begin{figure}[!htb]
\includegraphics[width=0.8\linewidth]{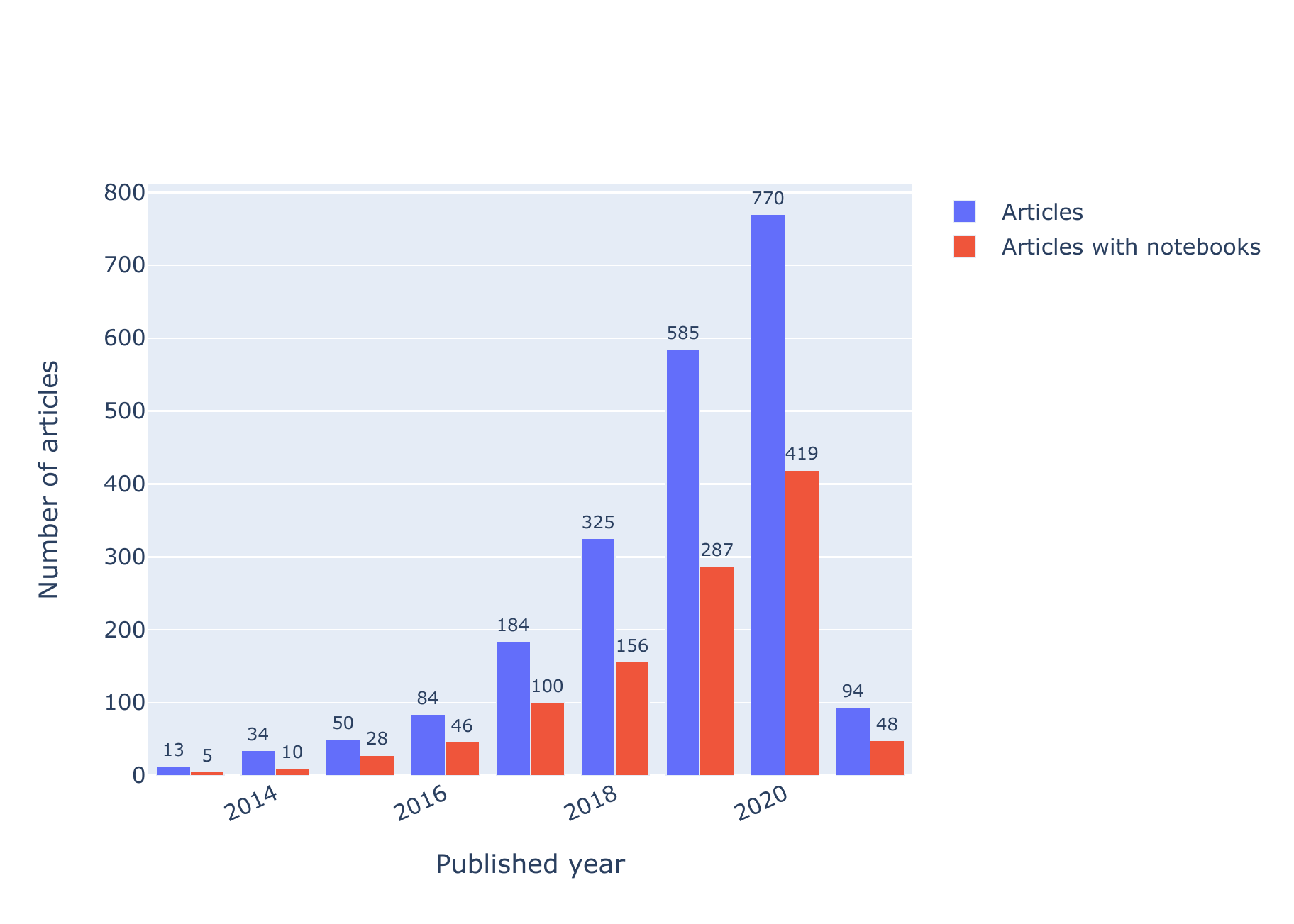}
\caption{Articles by number of GitHub repositories with at least one Jupyter notebook by year.}
\label{fig:Figure_timeline_articles_with_without_notebooks}
\end{figure}
\FIG{Figure_timeline_articles_with_without_notebooks} shows the maximum number of notebooks for articles published in the respective journal.
Among the top ten journals with notebooks, \textit{Nature Communications} had the maximum number of notebooks; however, it ranked fourth in terms of journals with repositories with notebooks.
\FIG{Figure_timeline_articles_with_without_notebooks} shows the timeline of the articles by the number of Github repositories with at least one Jupyter notebook.
This indicates a growing trend of articles with notebooks.\\
\begin{figure}[!htb]
\includegraphics[width=0.8\linewidth]{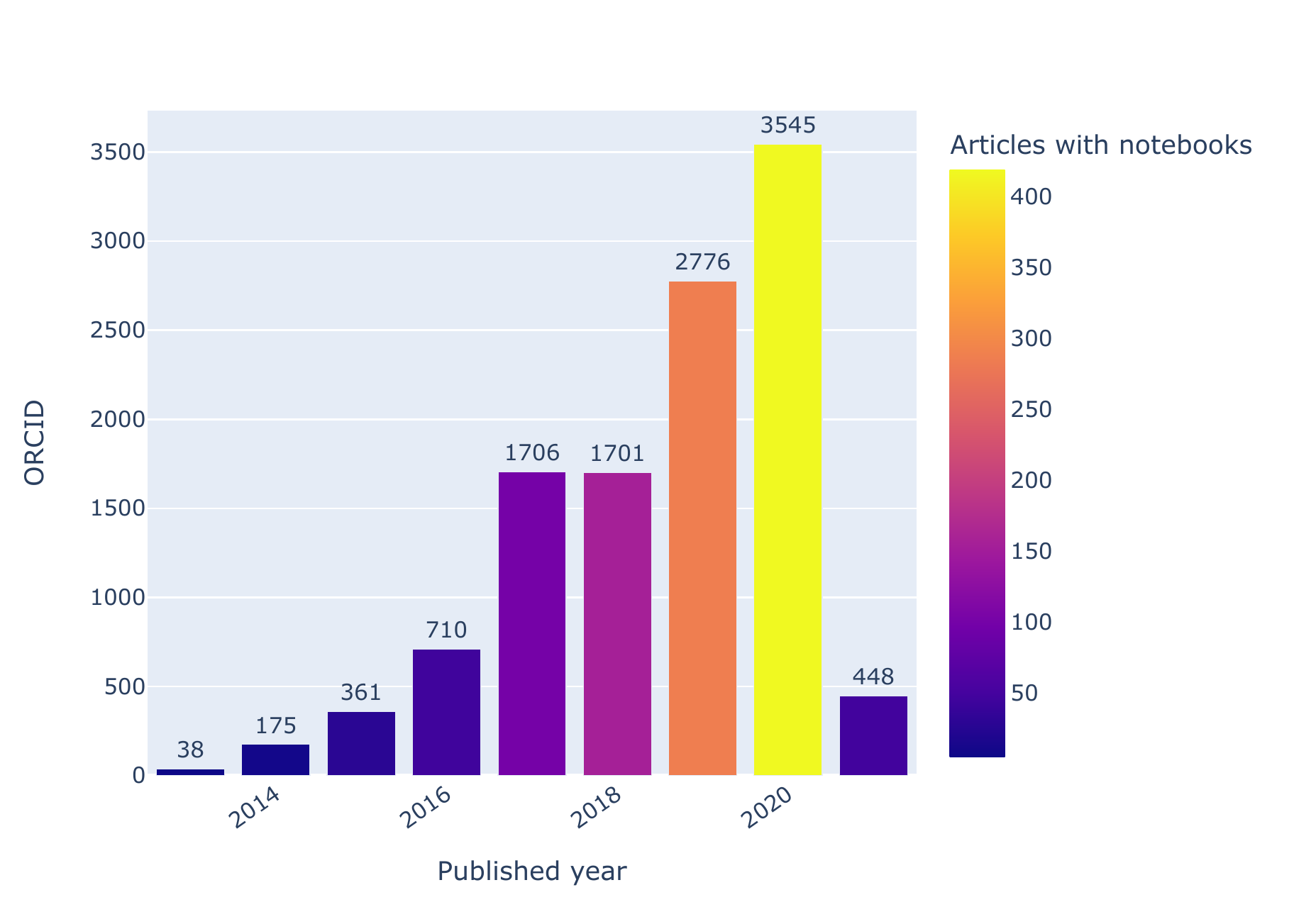}
\caption[ORCID usage in our collection.]{ORCID usage in our collection. Bars indicate the total number of ORCIDs found each year for authors of articles in our collection. Colors indicate the number of articles that year with Jupyter notebooks. Note that data for 2021 is incomplete, as only articles published by mid-February have been included.
}
\label{fig:Figure_timeline_articles_authors_with_orcid}
\end{figure}
In parallel to exploring trends related to Jupyter notebooks, we analyzed the uptake of ORCID identifiers\footnote{\url{https://orcid.org/}} over time in the collected journal articles with notebooks (\FIG{Figure_timeline_articles_authors_with_orcid}).
ORCID provides a persistent digital identifier to uniquely identify authors and contributors of scholarly articles. 
While iPython notebooks go back to 2001, the Jupyter notebooks with kernels for multiple languages became available in 2014, whereas ORCID was launched in 2012. Hence, both are relatively recent innovations in the scholarly communications ecosystem, and their respective uptake processes occur in parallel. 

There are in total 11594 authors in the 1419 publications. 
We have not performed any author disambiguation to distinguish unique authors in our corpus. However, such disambiguation is taking place at scale in Wikidata (see Discussion).
There are 2,720 (23.46\%) authors with ORCID and 8,874 (76.54\%) without.
In 2020, there were 3545 mentions of author ORCIDs and the highest number of articles (419) with notebooks.
The figure shows an increase in the usage of ORCID as well as articles with notebooks from 2016.

\subsection{Programming languages}
\label{sec:ProgrammingLanguages}
\begin{figure}[!htb]
\includegraphics[width=0.8\linewidth]{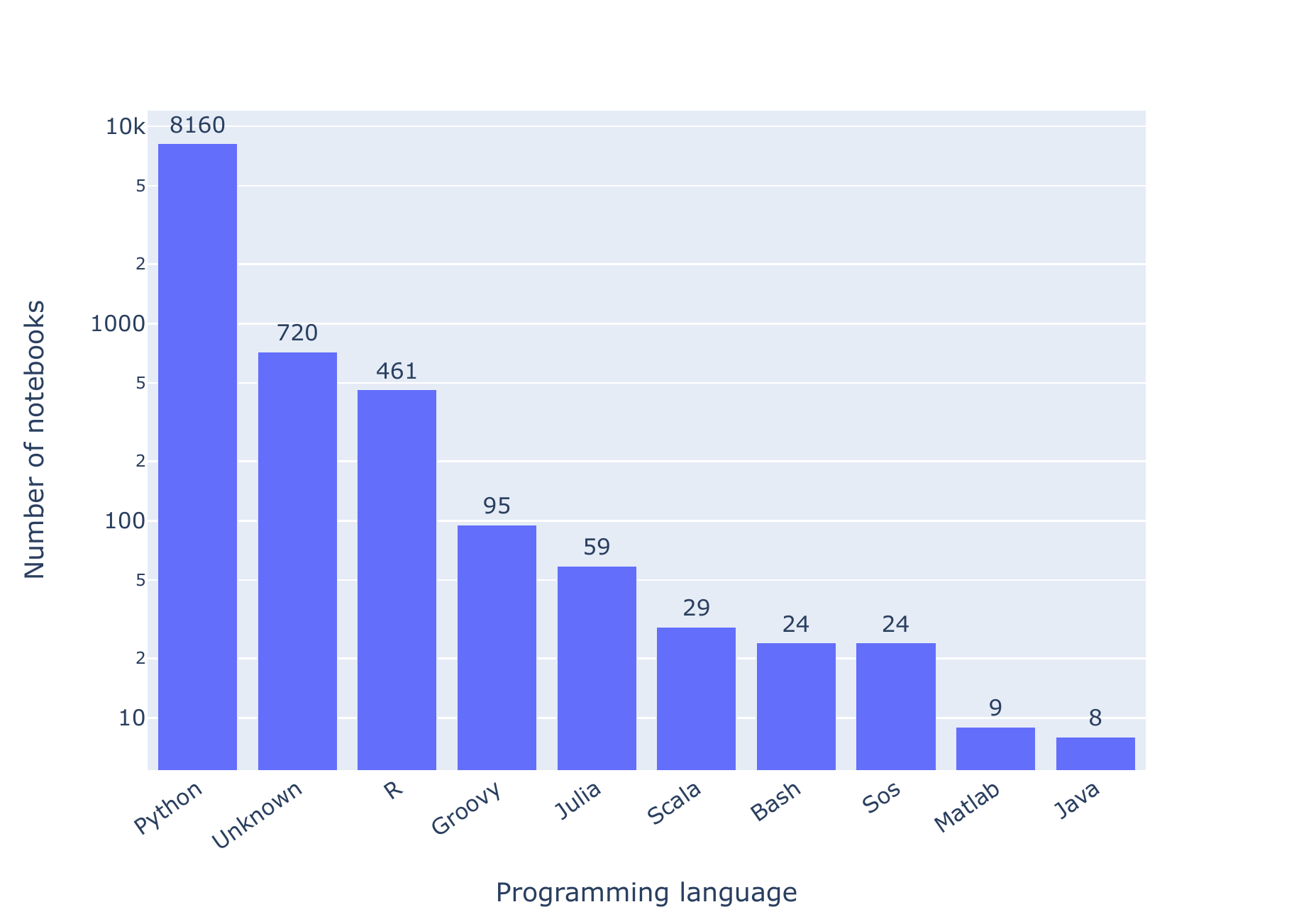}
\caption[Programming languages of the notebooks.]{Programming languages of the notebooks. ``Unknown'' means the language kernel used was not indicated in a standardized fashion.}
\label{fig:Figure_top_notebook_language}
\end{figure}
\begin{figure}[!htb]
\includegraphics[width=0.8\linewidth]{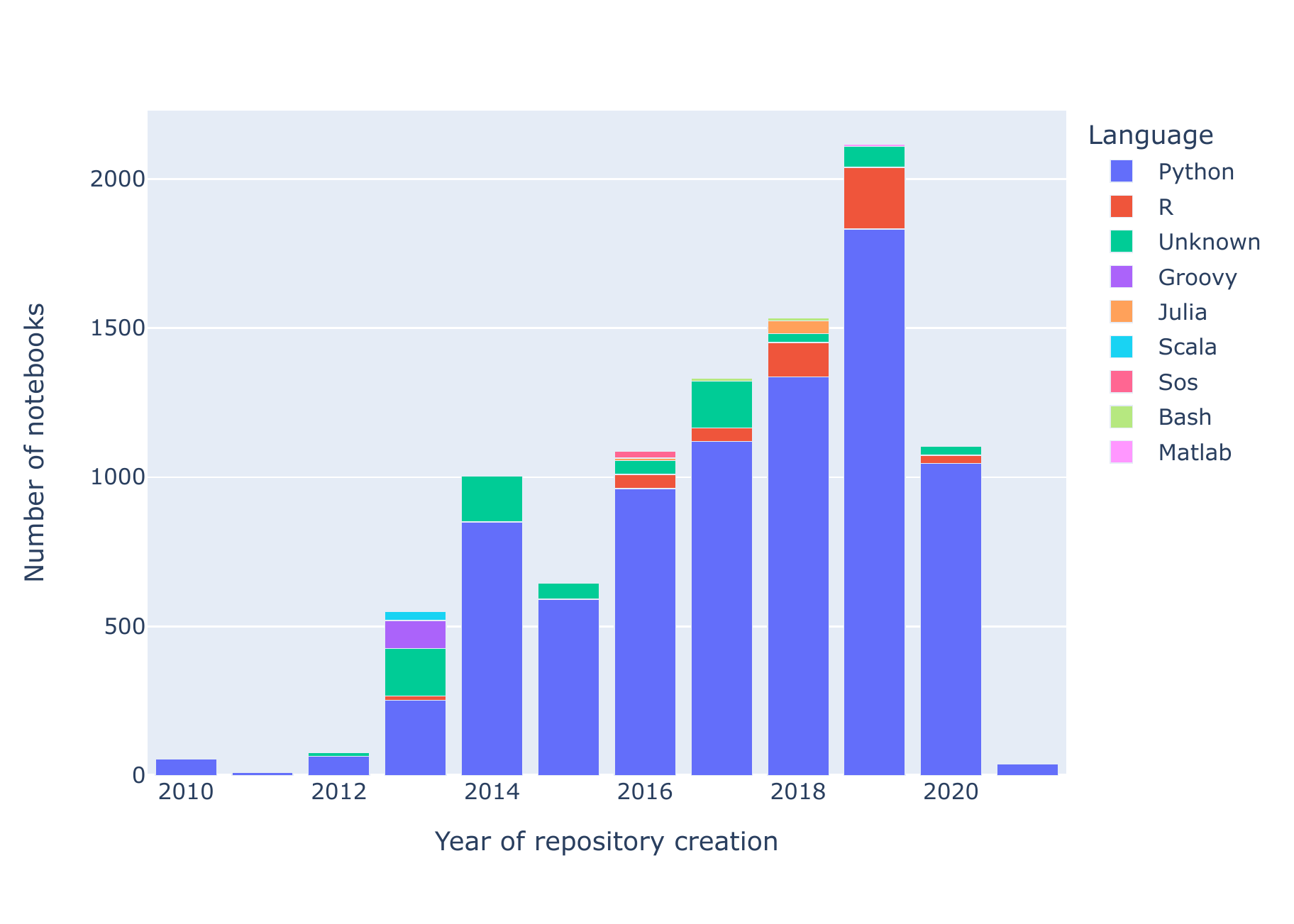}
\caption{Relative proportion of the most frequent programming languages used  in the notebooks per year.}
\label{fig:Figure_timeline_notebook_language_year}
\end{figure}
\FIG{Figure_top_notebook_language} and \FIG{Figure_timeline_notebook_language_year} show analyses of the programming languages used in the Jupyter notebooks present in the collected publications. 
\FIG{Figure_top_notebook_language} presents (using a log scale) the most common programming languages used in the notebooks.
Python (84.8\%) is the most common programming language, followed by unknown (7.5\%) and R (4.8\%).
Unknown notebooks are those which do not declare the programming language or its version in the notebook.
A total of 720 notebooks do not declare a programming language.
From the figure,  we can see that the Jupyter ecosystem is not just Python anymore, but Python is most prominent, and none of the other languages have overtaken the ``Unknown'' group, which is primarily due to early notebooks in which Python was hardcoded, or the language stated in some other non-standard fashion.
Jupyter Notebooks are also used for other languages like Bash, Matlab, and Java.
\FIG{Figure_timeline_notebook_language_year} shows the top programming languages used in notebooks based on the published year of the article.
There is a steady use of Python in Jupyter Notebooks. 
However, fewer notebooks have undeclared programming language in 2018 and 2020.
There is also an increase in the use of R in notebooks.
Authors can see the timeline of other programming languages in the analysis notebook provided in the repository.
\subsection{Versions}
\label{sec:Versions}
\begin{figure}[!htb]
\includegraphics[width=0.8\linewidth]{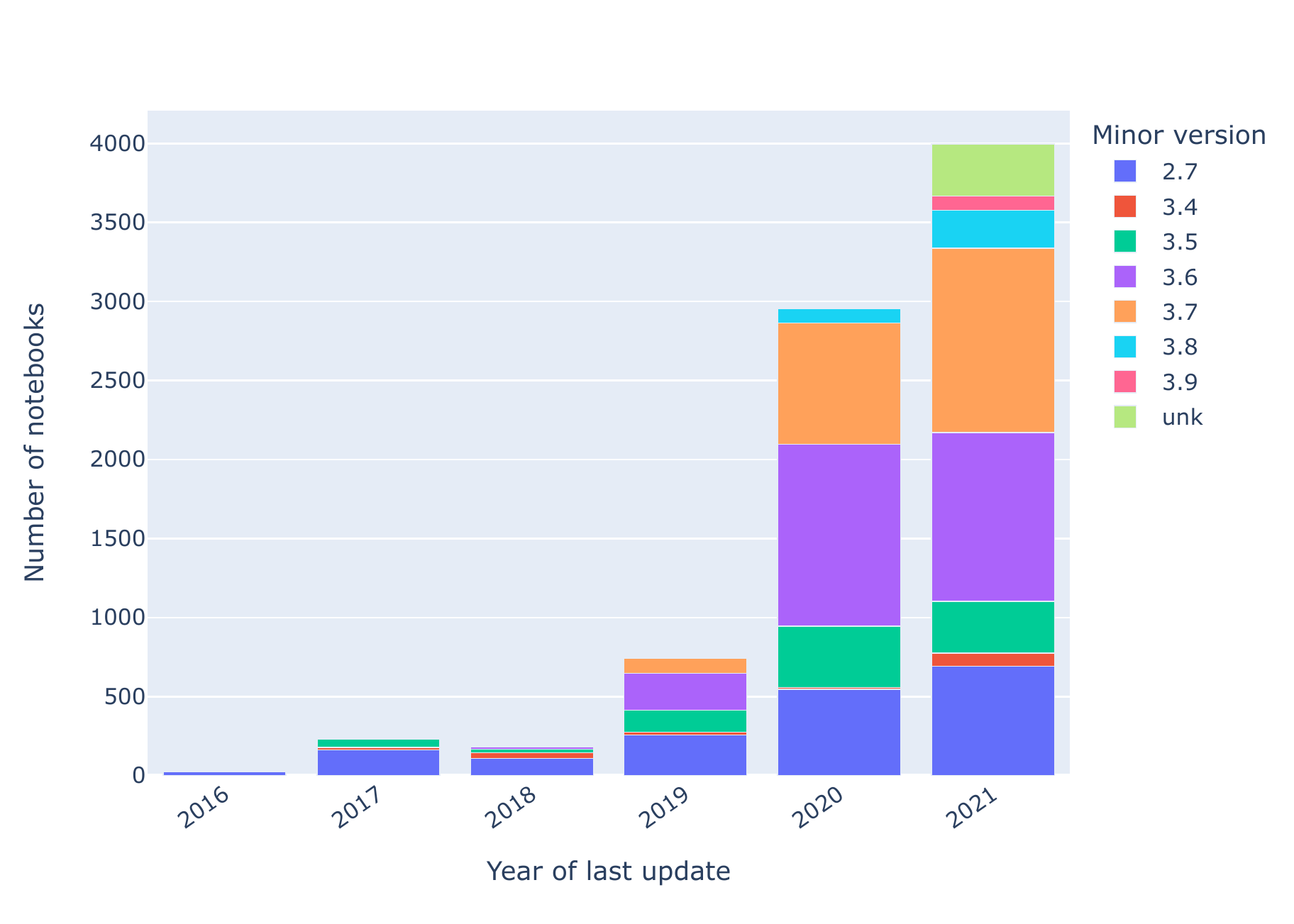}
\caption{Python notebooks by minor Python version by year of last commit to the GitHub repository.}
\label{fig:Figure_timeline_python_minor_version_by_repo_update}
\end{figure}
\begin{figure}[!htb]
\includegraphics[width=0.8\linewidth]{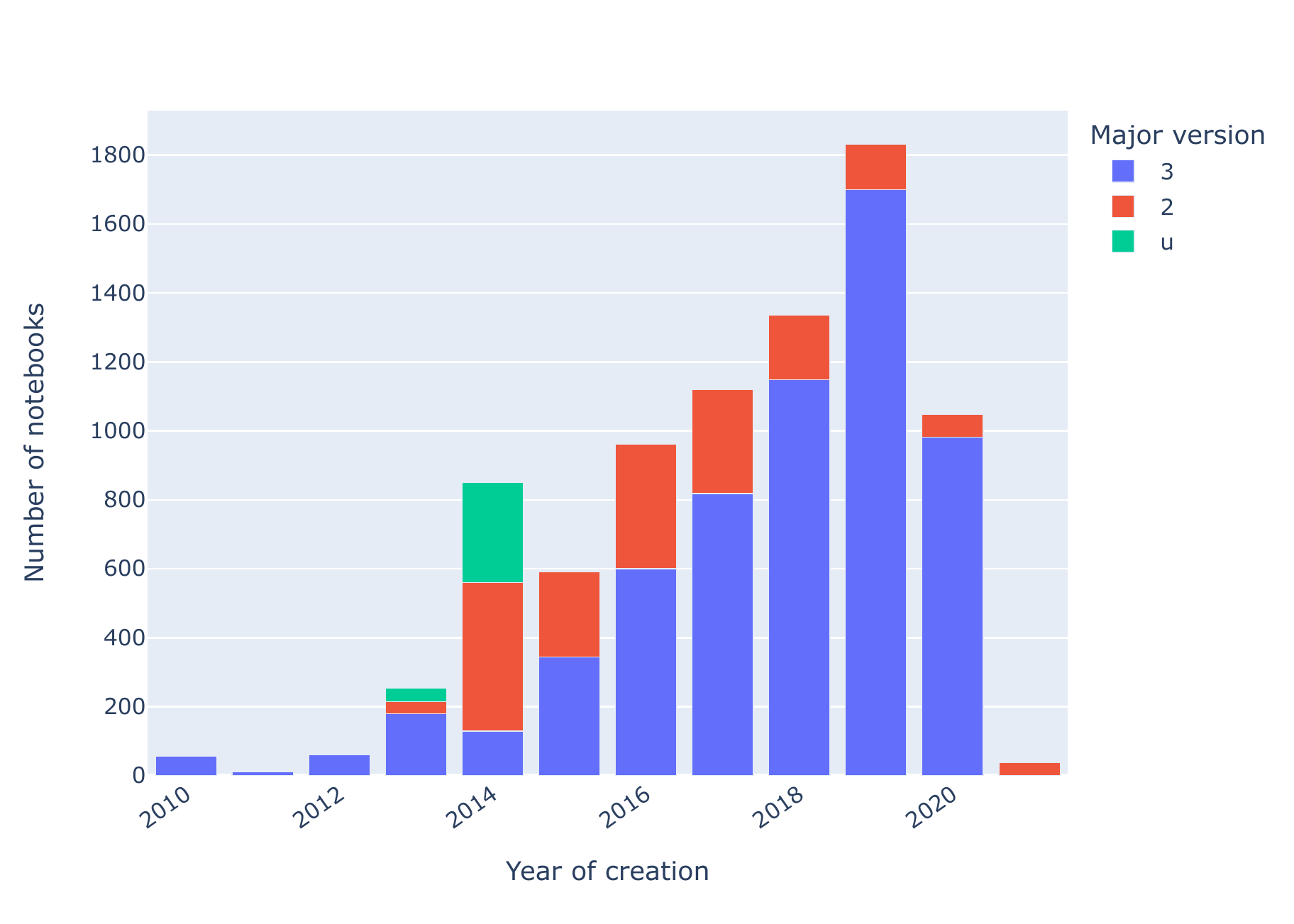}
\caption{Python notebooks by major Python version by year of first commit.}
\label{fig:Figure_timeline_python_major_version_by_repo_creation}
\end{figure}

\FIG{Figure_timeline_python_minor_version_by_repo_update} shows the Python version of notebooks based on the year in which the repository was last updated.
2471 notebooks have Python version 3.6, followed by 2031 notebooks with Python version 3.7.
Python version 3.6 and 3.7 are commonly used in recent years, followed by version 2.7.
There are also some python notebooks without any version declared.
6028 notebooks have Python major version 3, 1802 notebooks have Python major version 2, and 329 notebooks have an unknown Python version.
\subsection{Notebook structure}
\label{sec:NotebookStructure}
\begin{figure}[!htb]
    \centering
    \begin{subfigure}[t]{0.5\textwidth}
        \centering
        \includegraphics[width=\hsize]{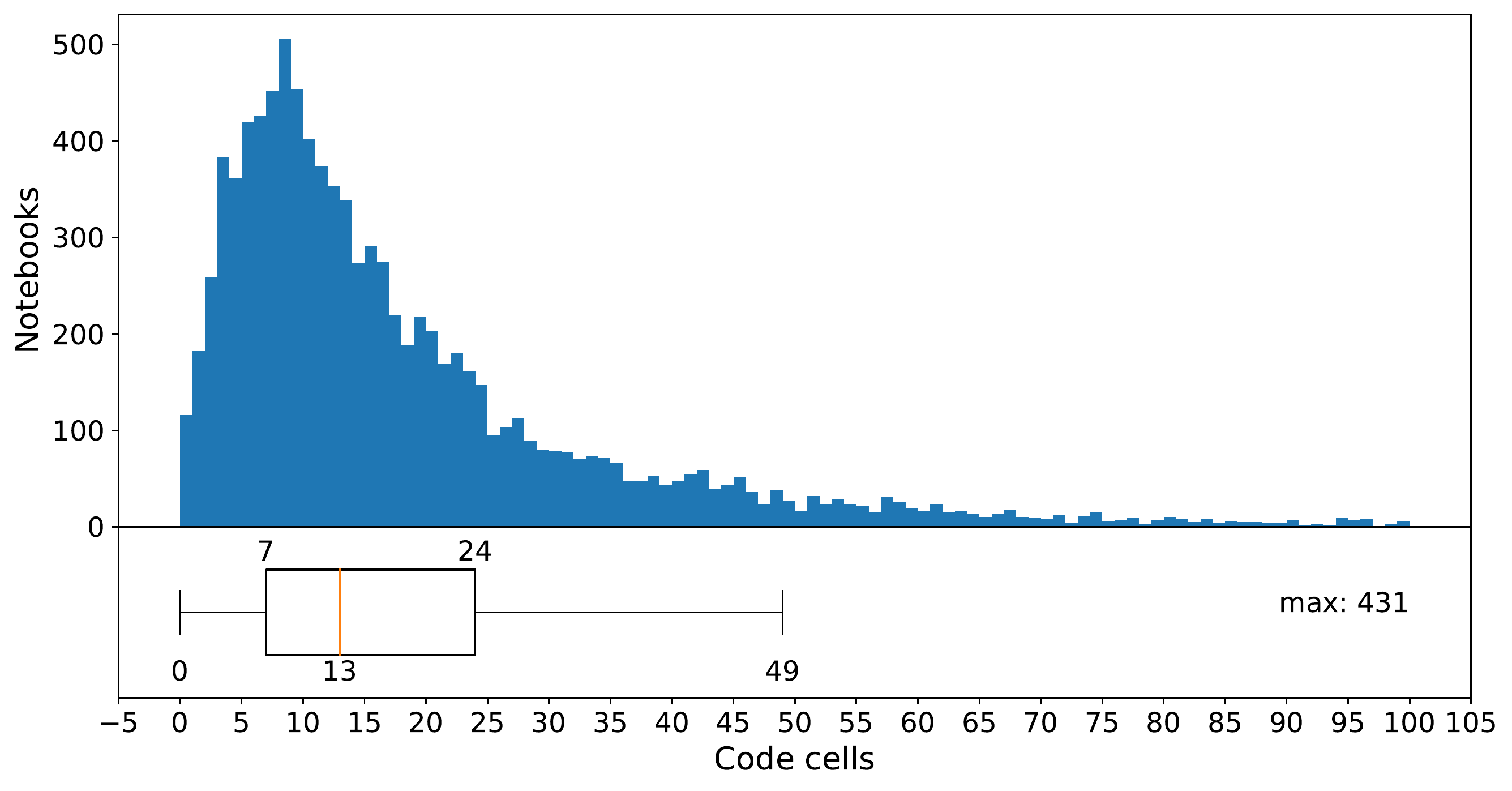}
    \caption{Distribution of the number of code cells across notebooks in our corpus. }
    \label{fig:Figure_f_code_cells}
    \end{subfigure}
    \hfill
    \begin{subfigure}[t]{0.45\textwidth}
        \centering
        \includegraphics[width=\hsize]{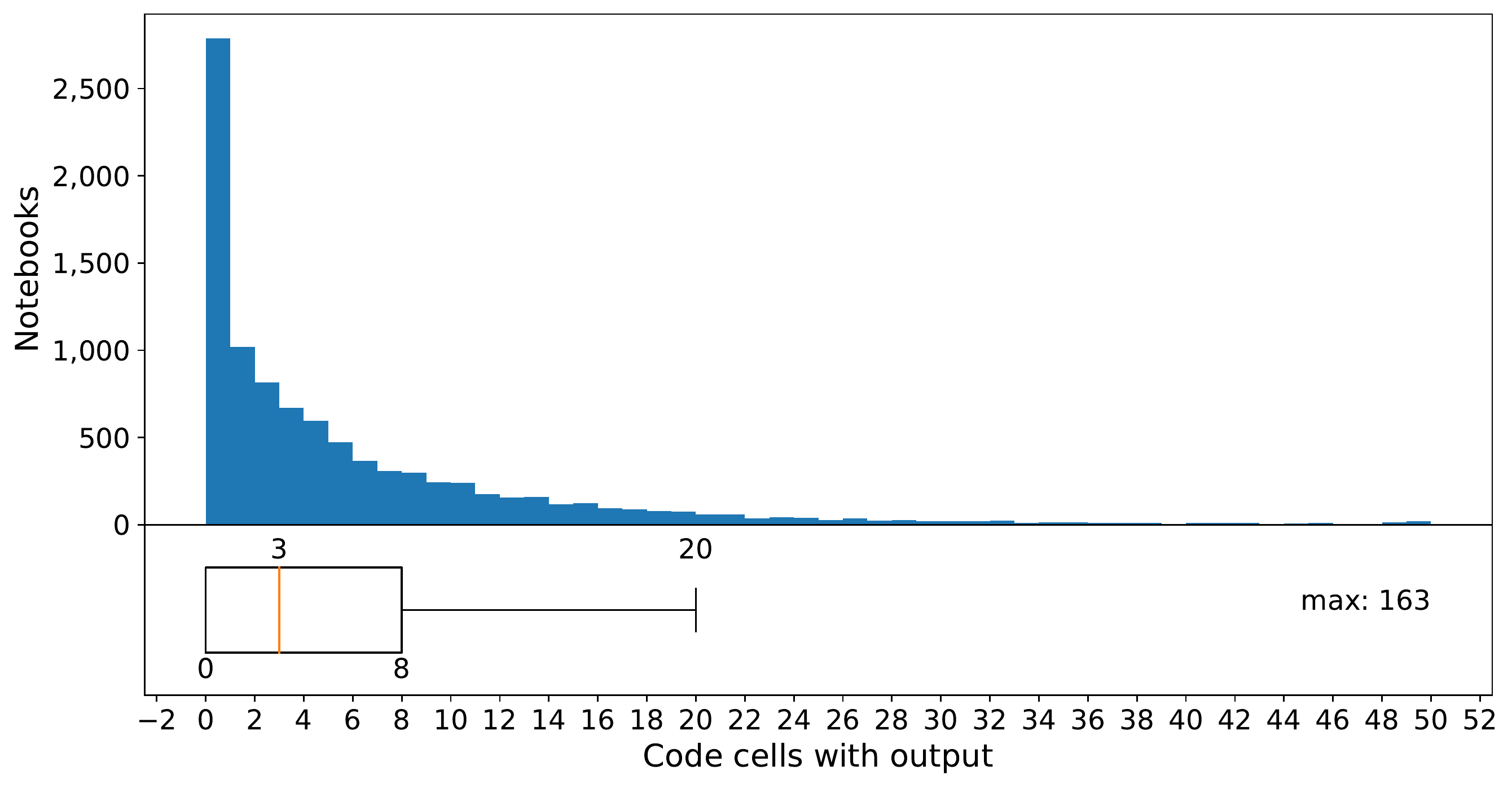}
\caption{Distribution of the number of code cells  with outputs across notebooks in our corpus.}
    \label{fig:Figure_f_code_cells_with_output}
    \end{subfigure}

    \vspace{1cm}
    \begin{subfigure}[t]{0.5\textwidth}
        \centering
        \includegraphics[width=\hsize]{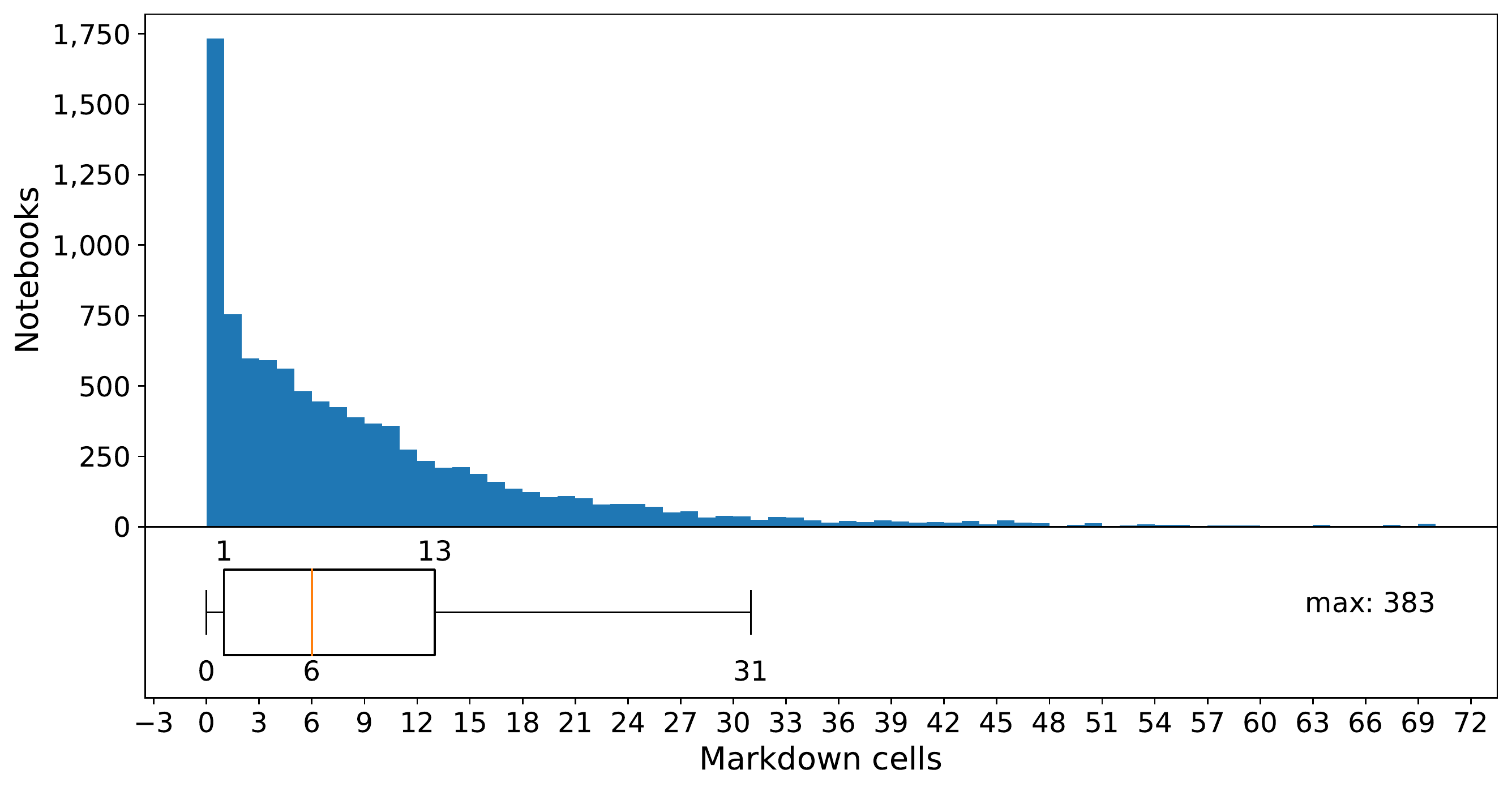}
        \caption{Distribution of the number of Markdown cells across notebooks in our corpus. }
        \label{fig:Figure_f_markdown}
    \end{subfigure}
    \hfill
    \begin{subfigure}[t]{0.45\textwidth}
        \centering
        \includegraphics[width=\hsize]{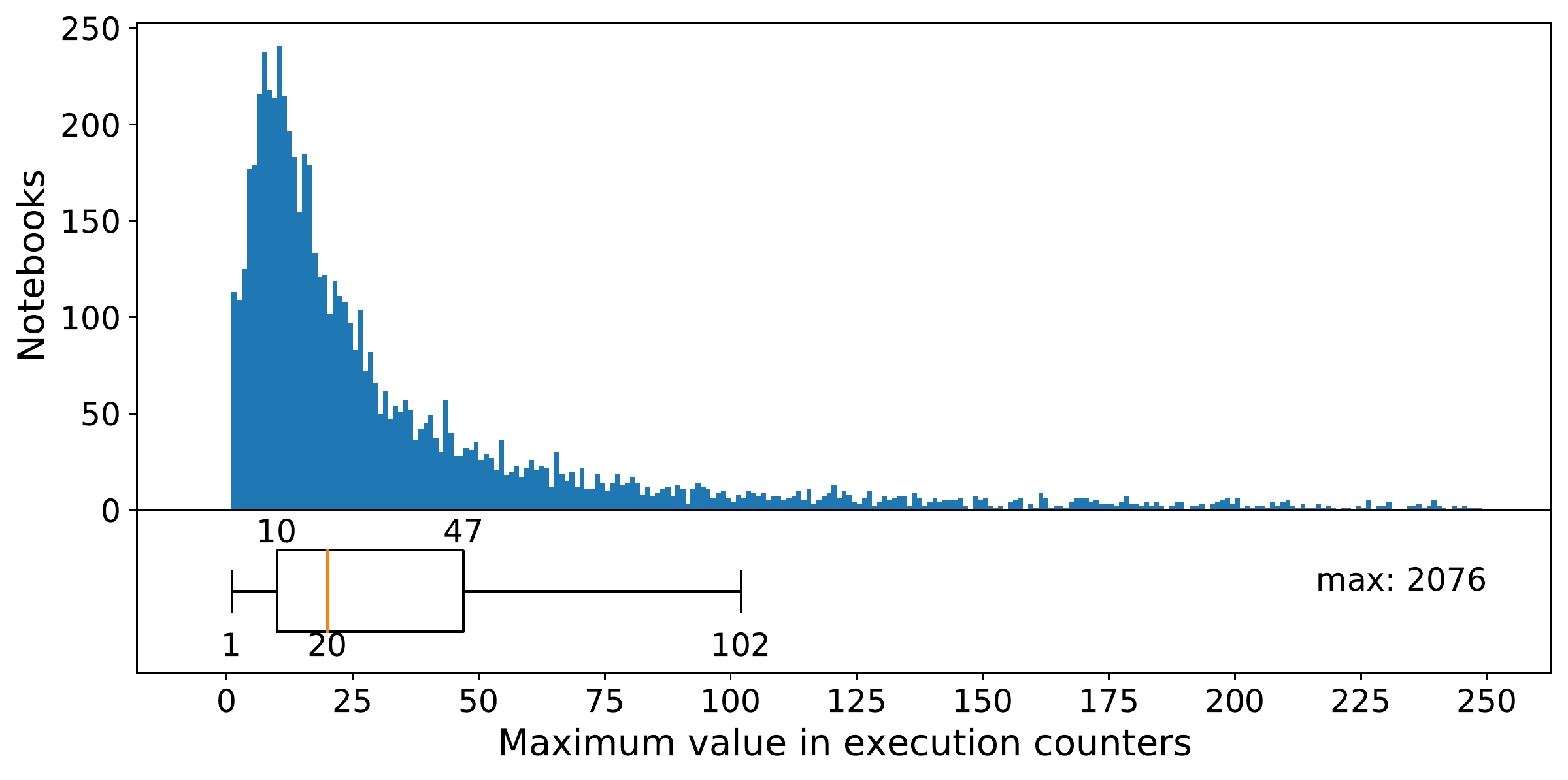}
        \caption{Distribution of the maximum execution count across notebooks in our corpus.}
        \label{fig:Figure_f_max_execution_count_full}
    \end{subfigure}
    \caption{Analysis of the notebook structure}
    \label{fig:Figure_notebook_structure_analysis}
\end{figure}

\FIG{Figure_notebook_structure_analysis} shows the statistics on the structure of notebooks.
Notebooks have a median of 20 cells and 13 code cells.
The average number of cells with outputs in notebooks found in our study is three, with zero being the least (\FIG{Figure_f_code_cells_with_output}).
The maximum number of cells, code cells, and cells with output seen in a notebook are 95, 431, and 163, respectively.
The maximum number of raw and empty cells seen in a notebook is 49 and 31, respectively.
Raw cells let the users write output directly and the kernel does not evaluate them.
The average number of markdown cells in notebooks is six, with the maximum being 383. 
6311 (65.77\%) of the notebooks have markdown cells, while 3284 (34.23\%) notebooks do not.
96.58\% of notebooks use English in the markdown cells; While 46.27\% notebooks use only English in the markdown cells.
In addition to English, French (11.76\%) and Danish (3.96\%) are the other popular natural languages used in the markdown.
In 1909 (30.25\%) notebooks, we could not detect the language in the markdown cells.
Further analysis of markdown cells shows that the average number of lines and words seen in markdown cells are 20 and 145, respectively.
Paragraphs and headers, the most commonly seen markdown elements, appear in 92.65\% and 81.81\% notebooks, respectively.
1,449 (17.76\%) notebooks do not have execution numbers and 6,710 (82.24\%) notebooks have execution numbers.
The maximum execution count seen in a notebook is 2076 (\FIG{Figure_f_max_execution_count_full}).
\subsection{Notebook naming}
\label{sec:NotebookNaming}
\begin{figure}[!htb]
\includegraphics[width=0.8\linewidth]{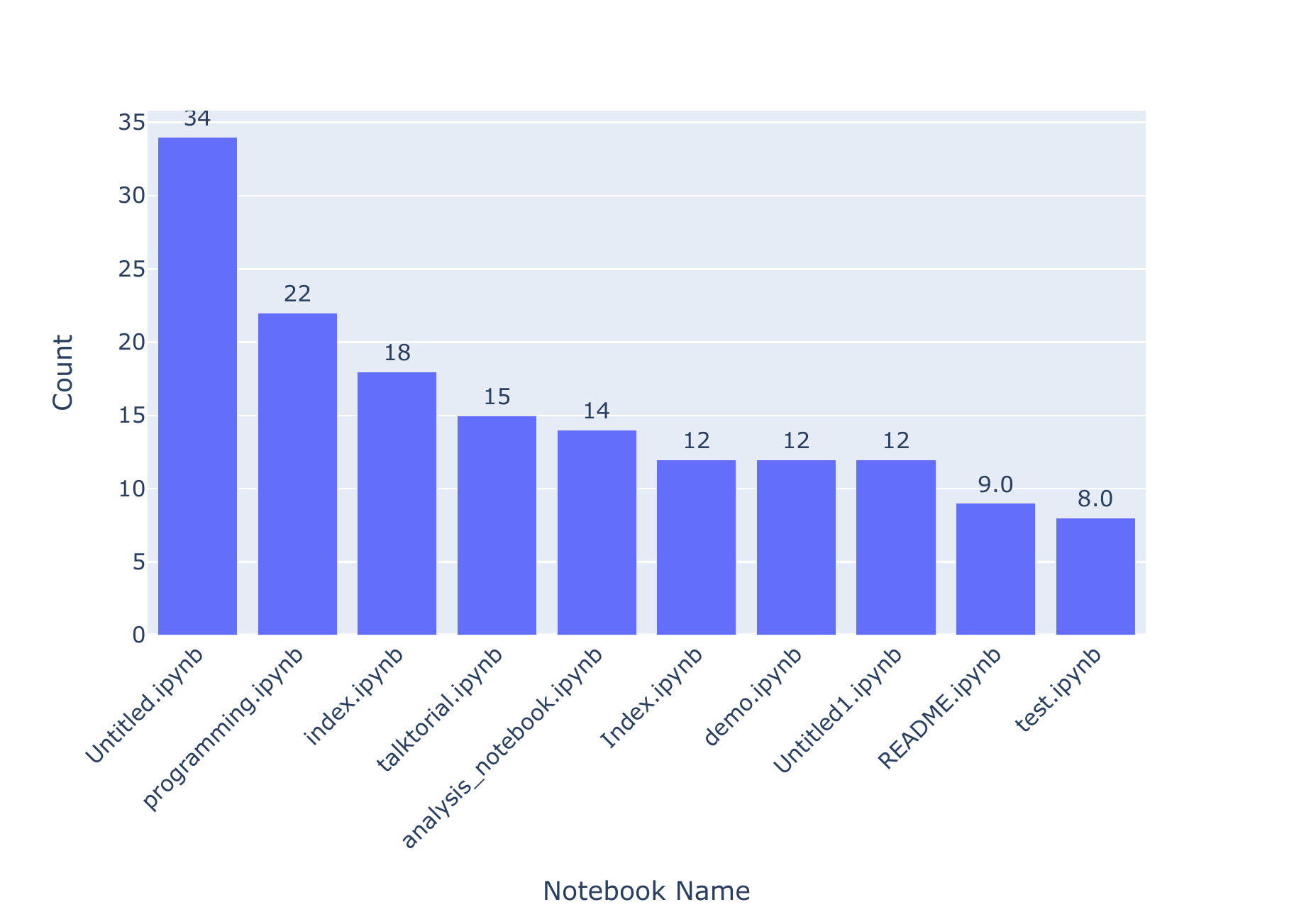}
\caption{Most frequent notebook titles. }
\label{fig:Figure_notebook_name_count}
\end{figure}
\begin{figure}[!htb]
\includegraphics[width=0.8\linewidth]{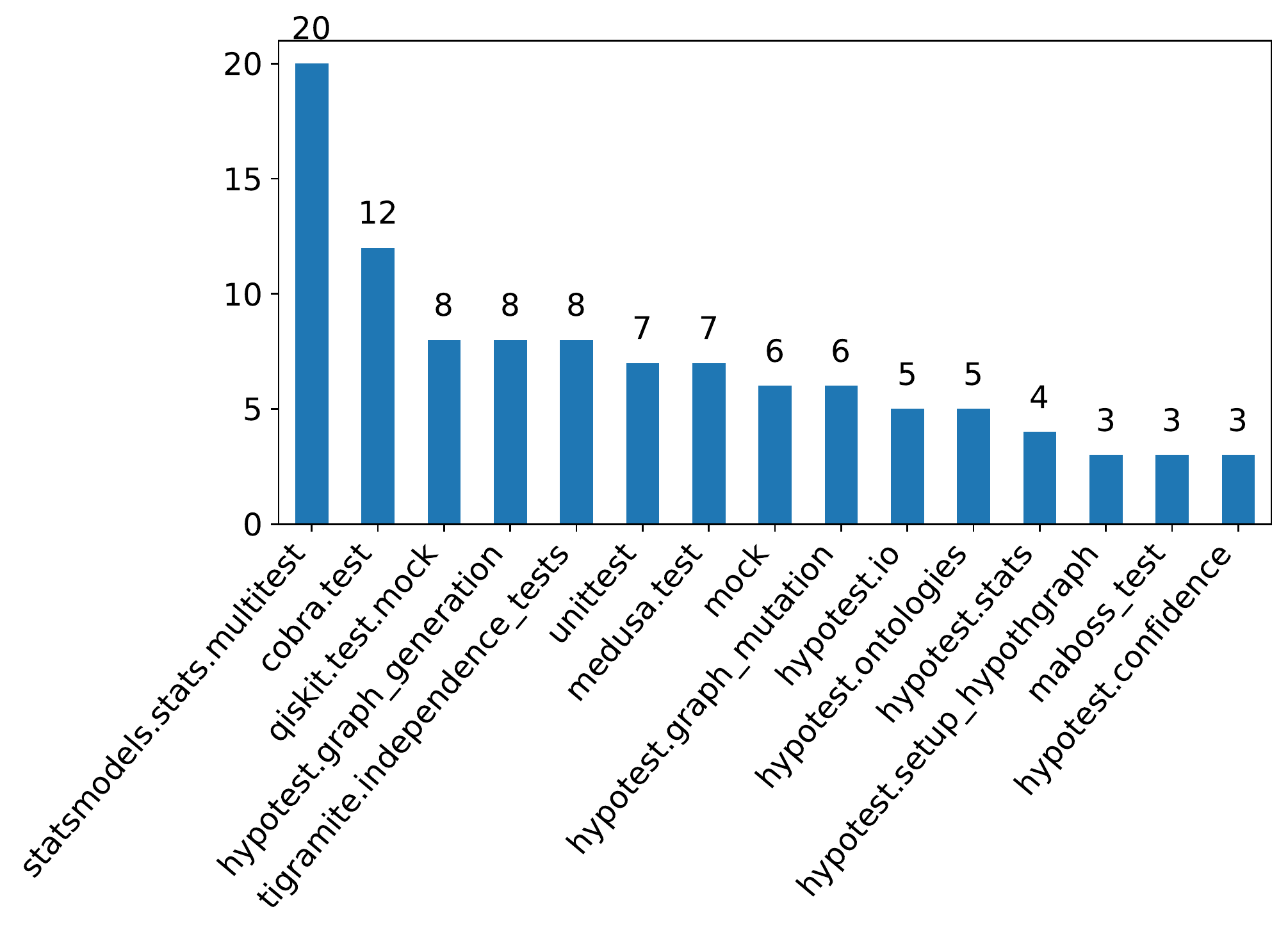}
\caption{Notebooks with the string \texttt{test}}
\label{fig:Figure_f_notebook_tests}
\end{figure}
\begin{figure}[!htb]
\includegraphics[width=0.8\linewidth]{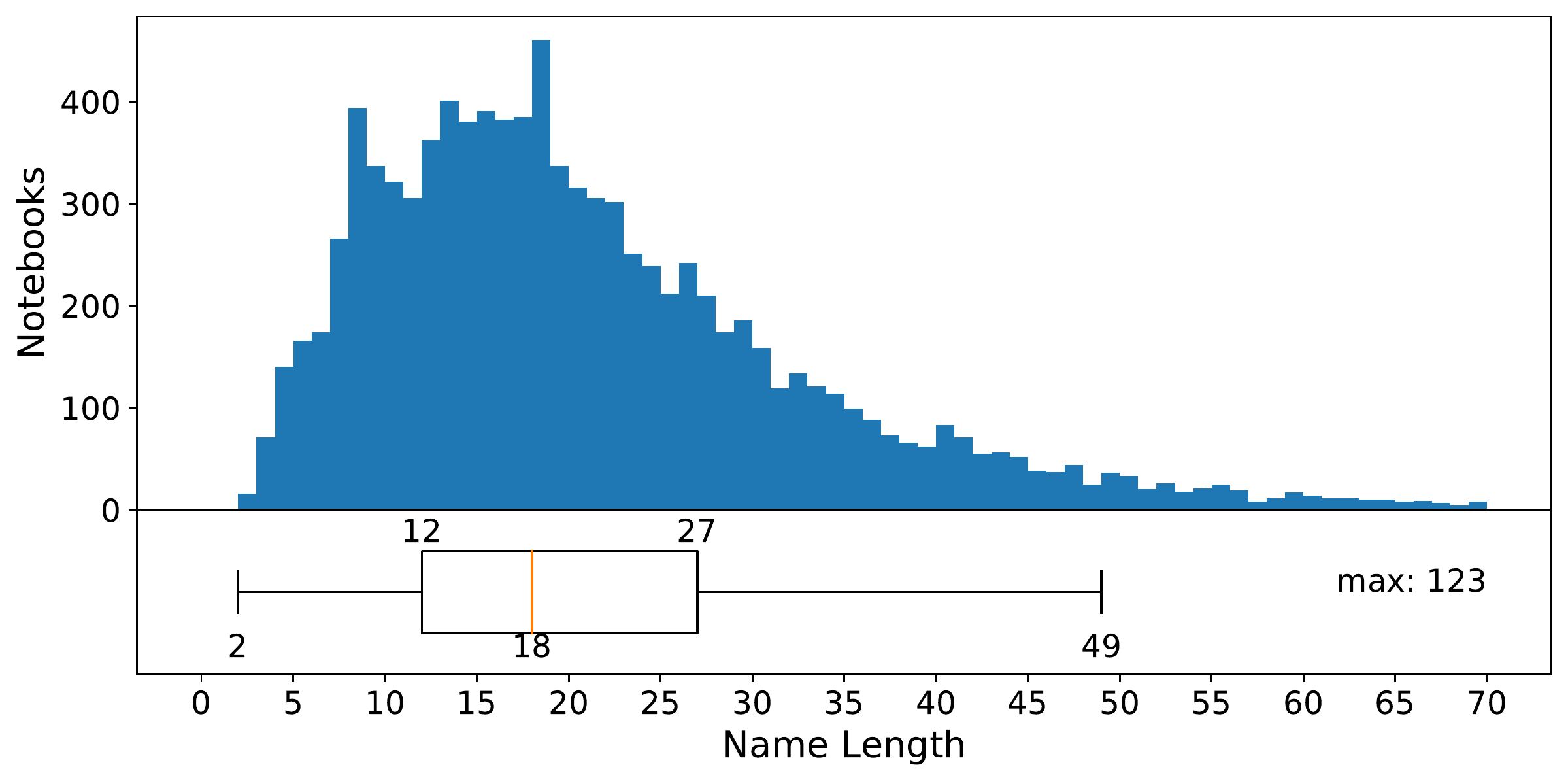}
\caption{Notebook title length }
\label{fig:Figure_f_notebook_name_length_full}
\end{figure}
\FIG{Figure_notebook_name_count} shows the most frequently used titles in notebooks from our collected data.
``Untitled'', ``programming'' and ``index'' are the three most common notebook names.
There are 63 (0.65\%) whose title is or starts with ``Untitled''.
There are 21 (0.22\%) notebooks that contain the name 'Copy'. 
We also see many notebooks with the string 'test' in their names (\FIG{Figure_f_notebook_tests}).
1,070 (11.12\%) notebooks have names that are not recommended by the POSIX fully portable filenames guide \citep{pimentel2019a}.
Only four notebooks have names that are disallowed in Windows.
There are no notebooks without a title (i.e., notebooks with just a '.ipynb' extension).
\FIG{Figure_f_notebook_name_length_full} shows the distribution of length of notebook title. 
The average length of the notebook title is 18 characters, with a maximum of 123 characters and a minimum of 2.

\subsection{Notebook modules}
\label{sec:NotebookModules}
\begin{figure}[!htb]
\includegraphics[width=0.5\textwidth]{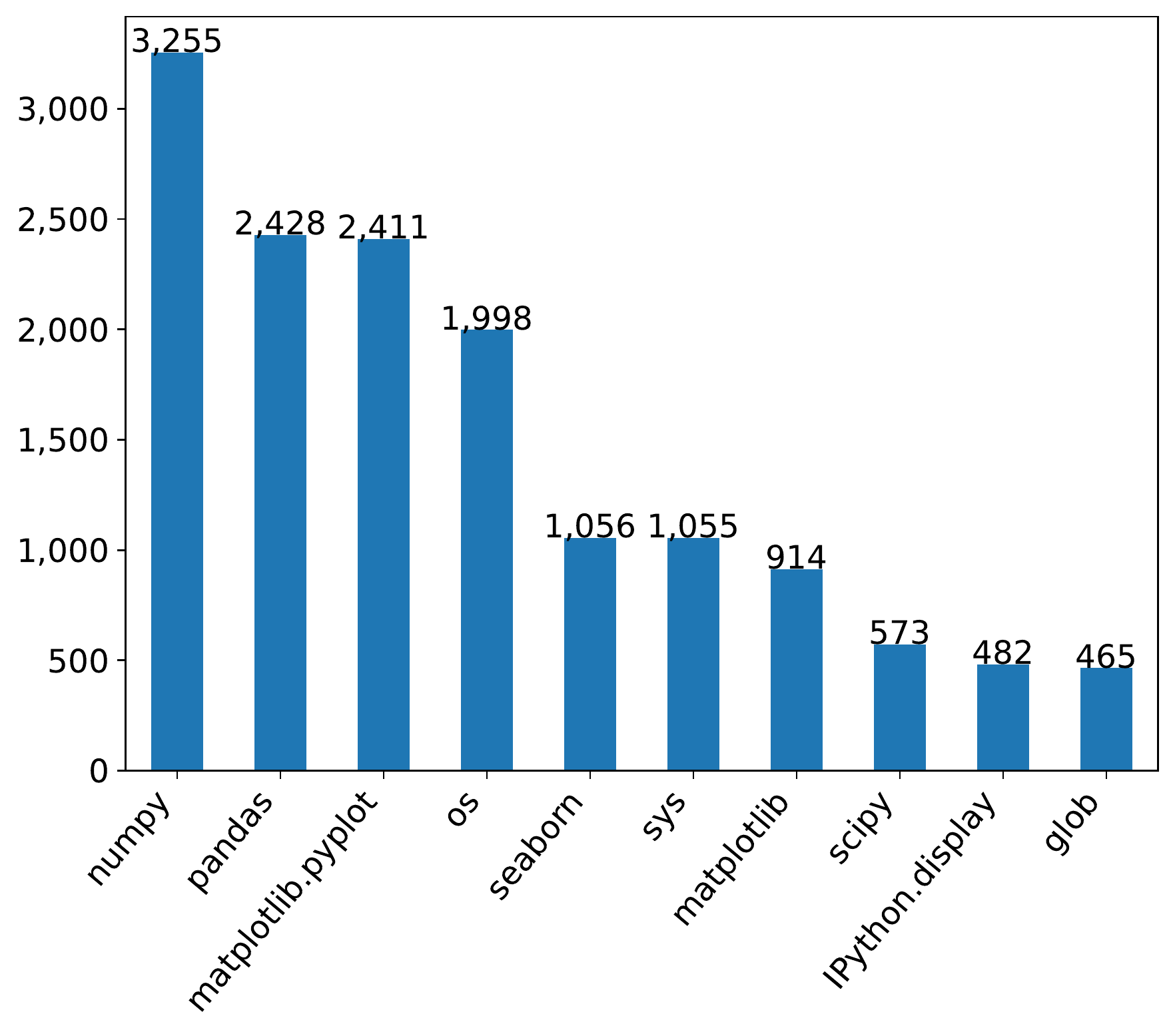}
\caption{Top Python modules declared in Jupyter Notebooks. }
\label{fig:Figure_f_notebook_module_full_import}
\end{figure}
\begin{figure}[!htb]
\includegraphics[width=0.5\textwidth]{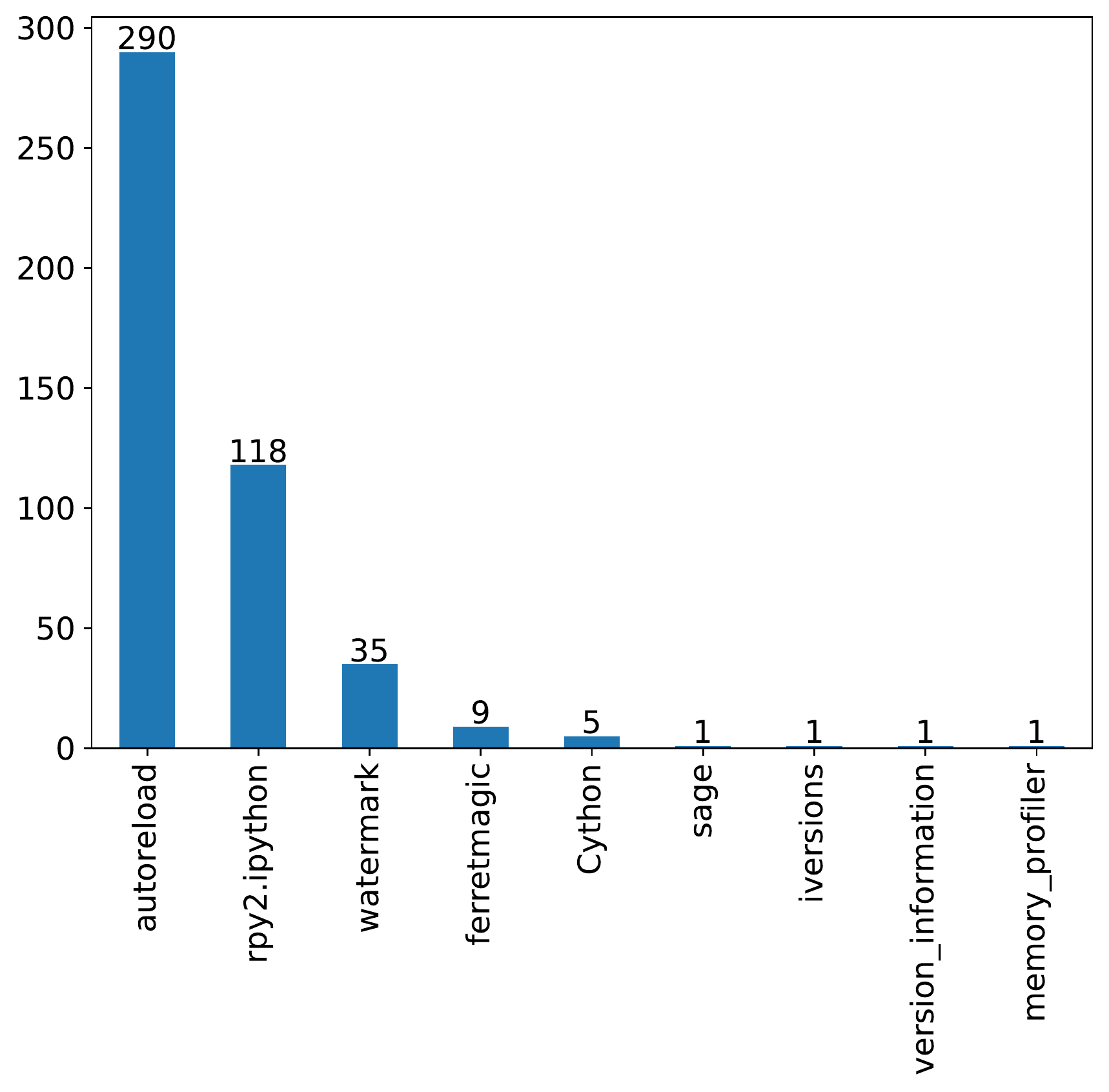}
\caption{Load extension modules in Jupyter Notebooks }
\label{fig:Figure_f_notebook_module_load_ext_full_import}
\end{figure}
\FIG{Figure_f_notebook_module_full_import} shows the analysis of modules declared in notebooks.
Using AST\footnote{\url{https://docs.python.org/3/library/ast.html}}, we analyzed the valid Python notebooks.
5,248 (69.06\%) notebooks had imports, of which 714 (9.40\%) had local imports, while 5,216 (68.64\%) had external modules.
Local imports denote the import of modules defined in the notebook repository's directory.
There are 1035 local and 38229 external modules declared in the collected Python notebooks.
\FIG{Figure_f_notebook_module_full_import} shows the top ten commonly used Python modules declared in the notebooks.
The most used modules are \textit{numpy} (3255), \textit{pandas} (2428), and \textit{matplotlib.pyplot} (2411). 
These are widely used modules for data manipulation, analytics, and visualizations.

\subsection{Notebook dependencies}
\label{sec:NotebookDependencies}
\begin{figure}[!htb]
    \centering
    \begin{subfigure}[t]{0.45\textwidth}
        \centering
        \includegraphics[width=0.8\linewidth]{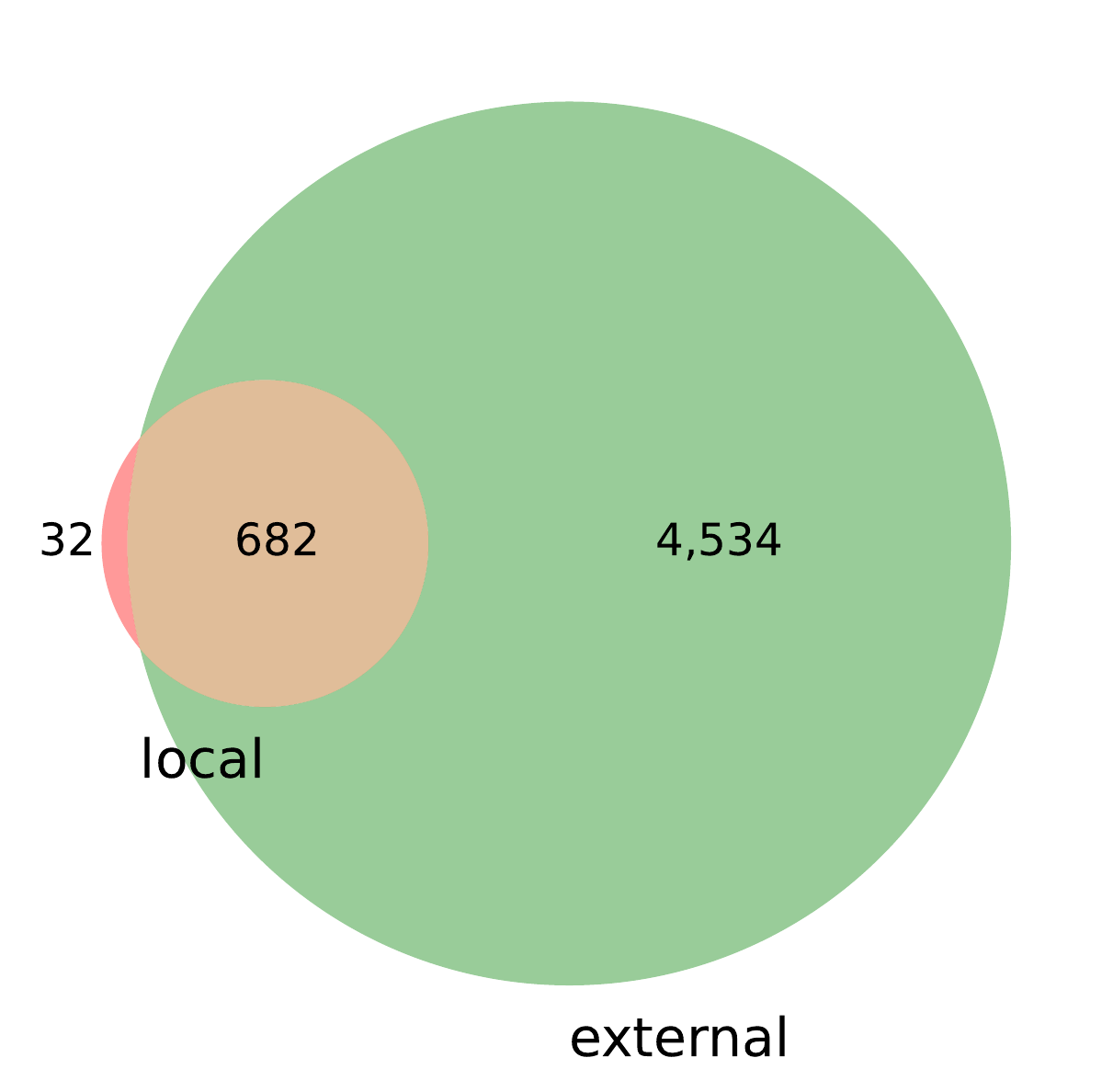} 
        \caption{External versus local modules declared in Jupyter notebooks.} \label{fig:Figure_f_notebook_module_external_local}
    \end{subfigure}
    \hfill
    \begin{subfigure}[t]{0.45\textwidth}
        \centering
        \includegraphics[width=\linewidth]{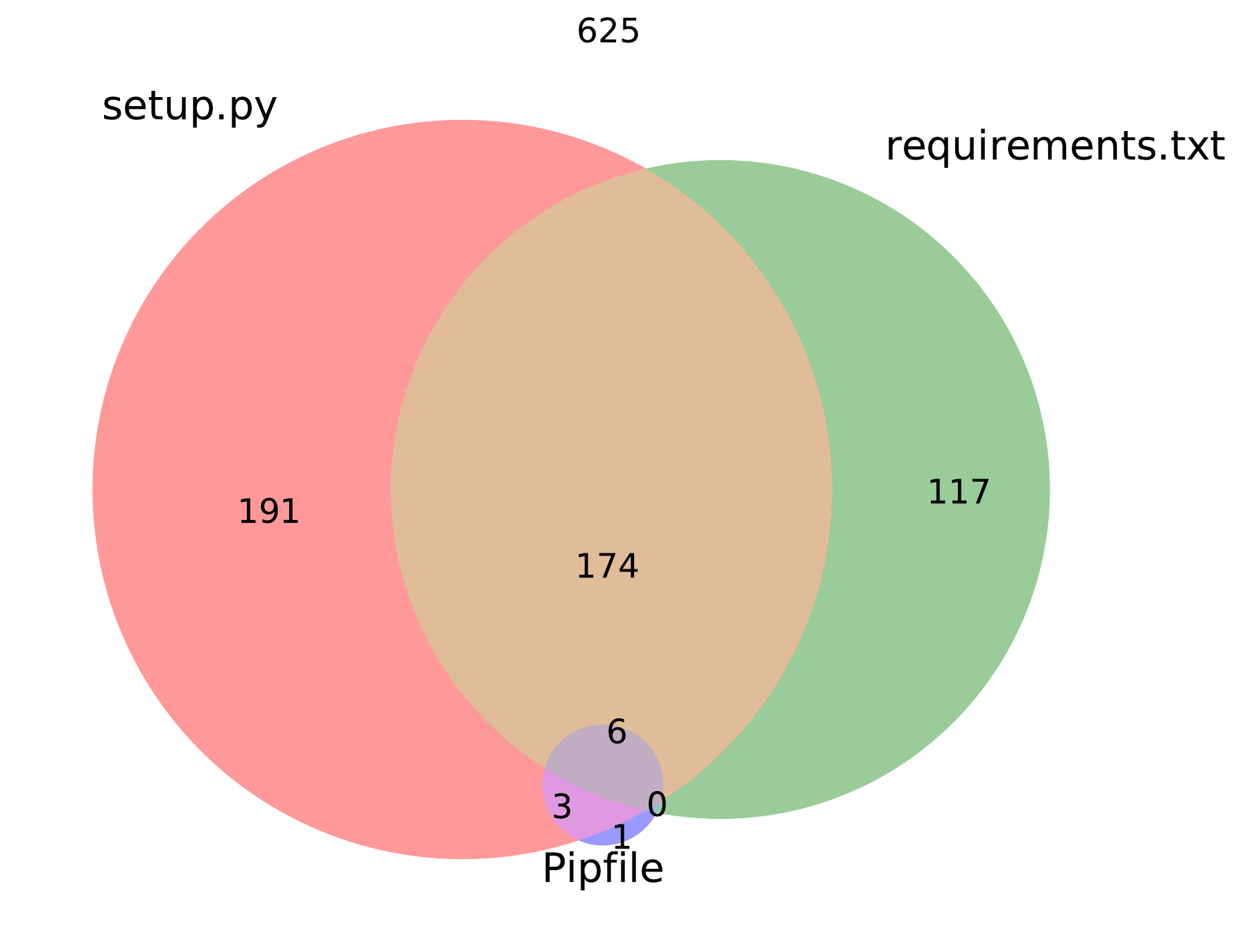} 
        \caption{Repositories with dependencies.} \label{fig:Figure_a_repository_dependencies_x3}
    \end{subfigure}

    \vspace{1cm}
    \begin{subfigure}[t]{0.5\textwidth}
    \centering
        \includegraphics[width=\linewidth]{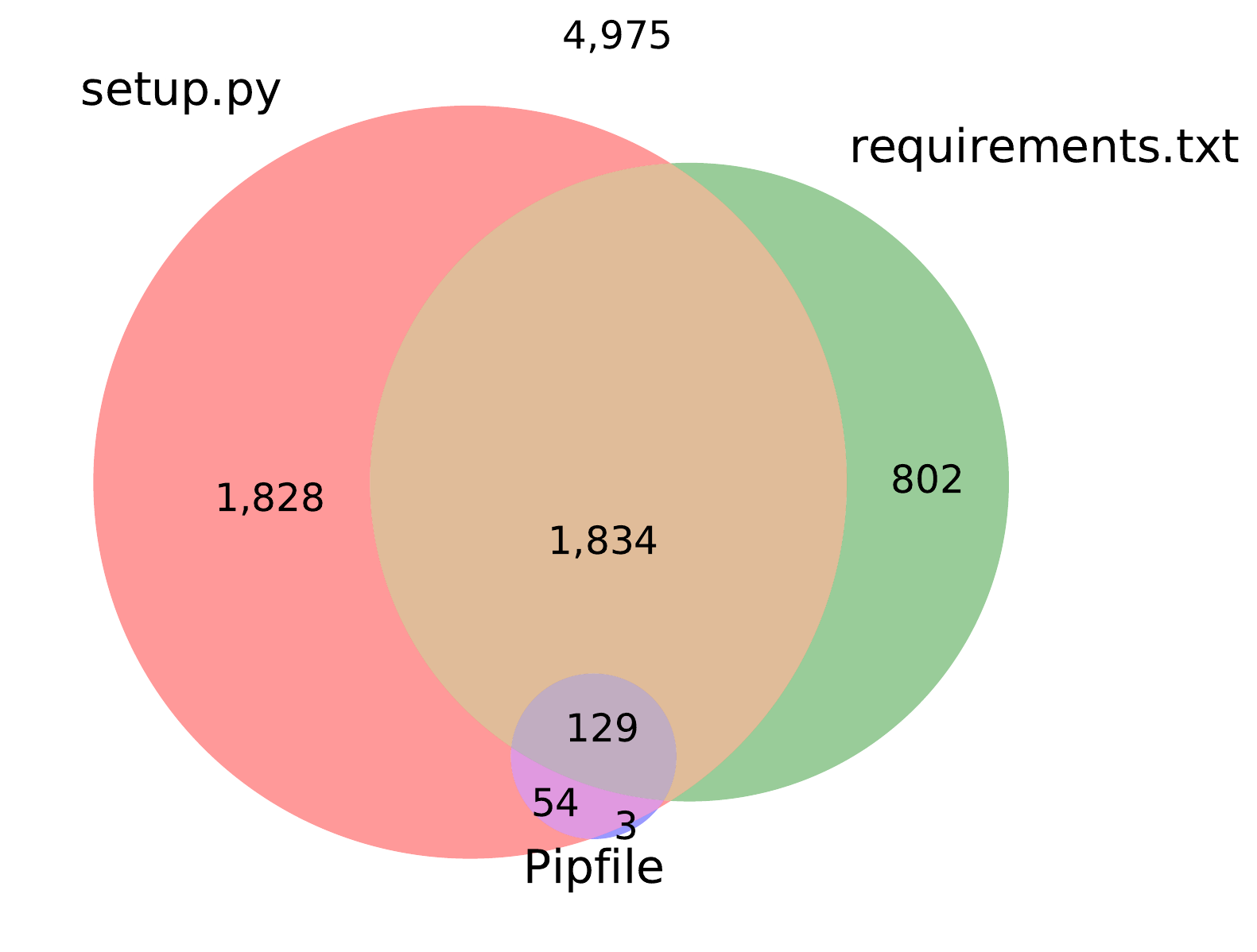} 
        \caption{Notebooks with dependencies.} \label{fig:Figure_a_notebook_dependencies_x3}
    \end{subfigure}
    \caption[Dependencies of Juypter Notebooks and GitHub repositories.]{Dependencies of Juypter Notebooks and GitHub repositories. In (\subref{fig:Figure_f_notebook_module_external_local}), the notebooks depending on external modules (green) are plotted against notebooks depending on local modules (red) and notebooks that had both (brown). In (\subref{fig:Figure_a_repository_dependencies_x3}) and  (\subref{fig:Figure_a_notebook_dependencies_x3}), GitHub repositories and Jupyter notebooks are shown as to whether they declared their dependencies via any combination of \texttt{setup.py} (red),  \texttt{requirements.txt} (green) or a \texttt{pipfile} (pink).}
    \label{fig:Figure_dependencies}
\end{figure}
\FIG{Figure_dependencies} shows the analysis of the declared dependencies of GitHub repositories and notebooks. 
4650 (48.31\%) of notebooks 
belong to repositories which have declared dependencies using \textit{setup.py}, \textit{requirements.txt}, or \textit{pipfile}.
There are 492 repositories with declared dependencies (\FIG{Figure_a_repository_dependencies_x3}).
There are 194 repositories with \textit{setup.py} file, 117 repositories with \textit{requirements.txt} file. 
180 repositories have both \textit{setup.py} and \textit{requirements.txt} file.
Only 10 repositories are with \textit{pipfile} (0.90\%).
In our study, 3845 (39.95\%) of notebooks use \textit{setup.py} file, 2765 (28.73\%) notebooks use \textit{requirements.txt} and only 186 (1.93\%) notebooks use \textit{pipfile}.

\subsection{Notebook Reproducibility}
\label{sec:NotebookReproducibility}
In our reproducibility study, we executed 4169 (43.45\%) Python notebooks.
The dependencies of the notebooks, as mentioned in their respective repositories, were installed in conda environments.
But, dependencies of 1,485 (35.62\%) notebooks failed to install.
None of the files were malformed with wrong syntax or conflicting dependencies.
We did not find any missing files that required other requirement files which were unavailable or files that needed external tools.
Hence, the reason for the failed installed error is unknown. 
We attempted to execute 2,684 (64.38\%) notebooks for the reproducibility study after successfully installing all the requirements.
However, many notebooks failed to execute even after installing all the requirements successfully. 

\subsection{Exceptions}
\label{sec:Exceptions}
\begin{figure}[!htb]
\includegraphics[width=0.8\linewidth]{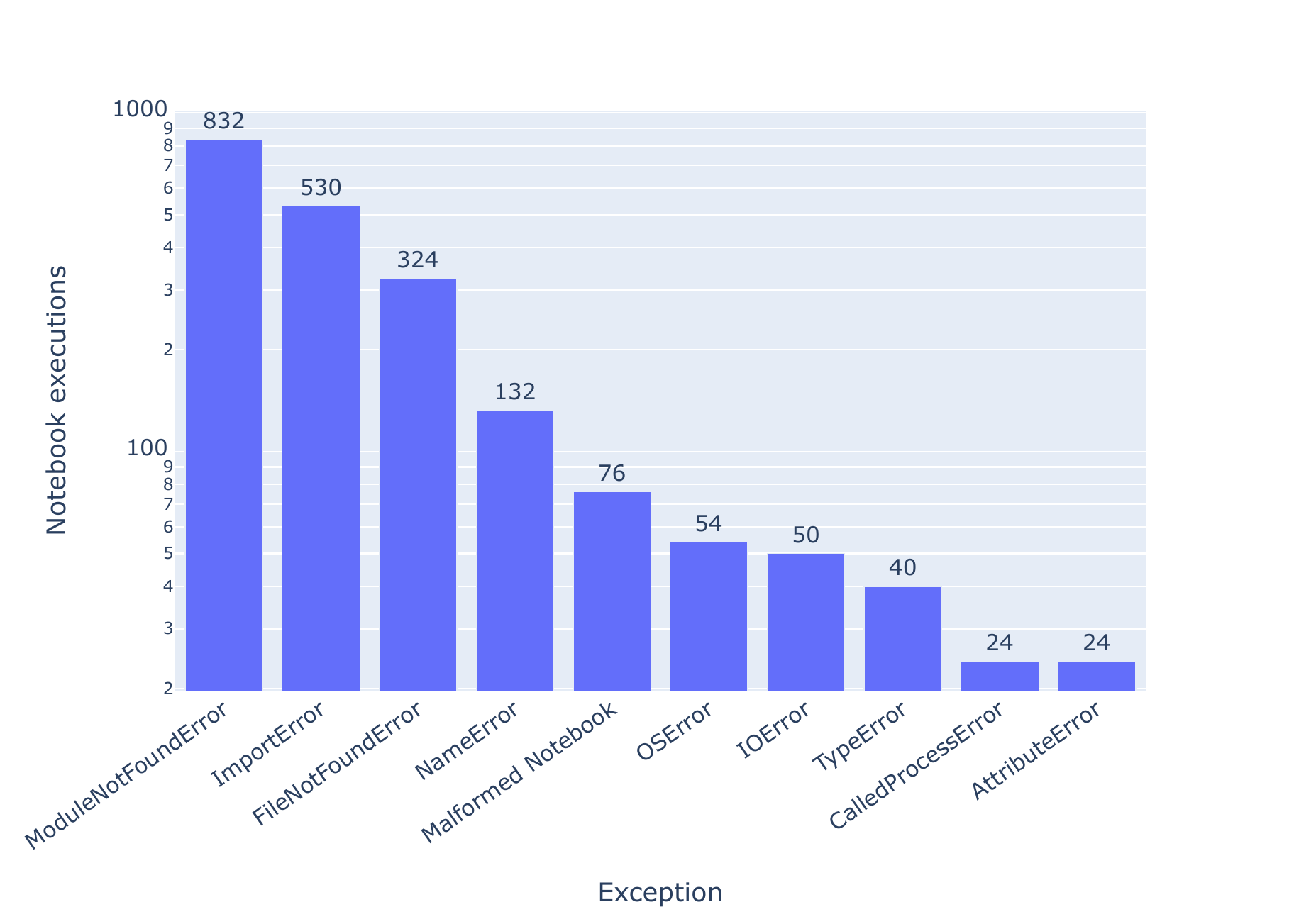}
\caption{Exceptions occurring in Jupyter Notebooks.}
\label{fig:Figure_top_exception_by_reason}
\end{figure}
\begin{figure}[!htb]
\includegraphics[width=0.8\linewidth]{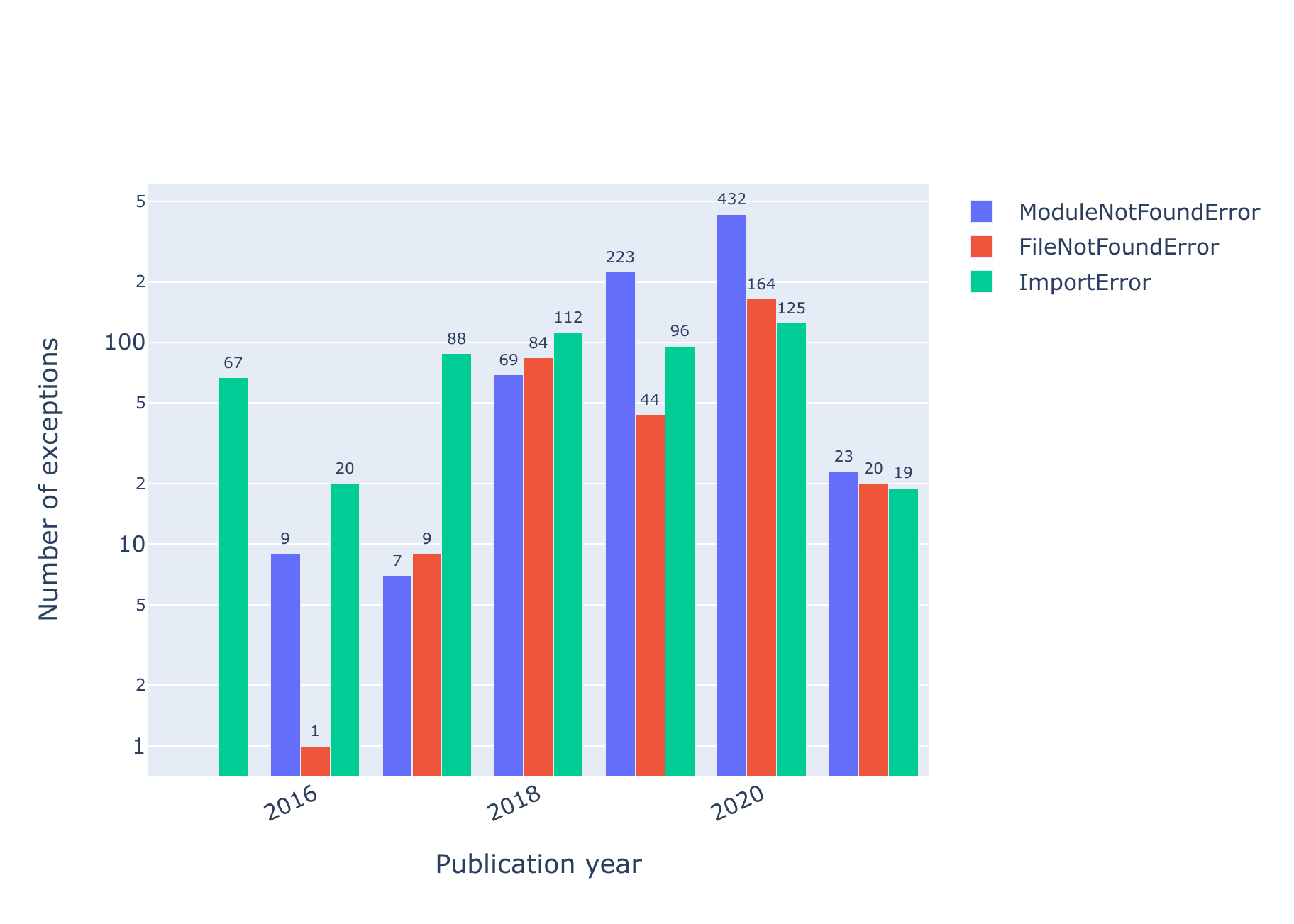}
\caption{ModuleNotFoundError, ImportError and FileNotFoundError exceptions by year of publication. }
\label{fig:Figure_timeline_exceptions_per_article}
\end{figure}
2,265 (84.39\%) notebooks resulted in exceptions due to several reasons.
\FIG{Figure_top_exception_by_reason} shows the top ten exceptions that occurred while executing the notebooks.
\textit{ModuleNotFoundError}, \textit{ImportError}, and \textit{FileNotFoundError} are the most common reason that resulted in failure of execution in notebooks.
1,362 (32.67\%) of the executions failed because of \textit{ModuleNotFoundError} and \textit{ImportError} exceptions.
\textit{ModuleNotFoundError} exception occurs when a Python module used by the notebook could not be found.
\textit{ImportError} exception occurs when a Python module used by the notebook could not be imported.
These two errors occur mainly due to missing dependencies.
132 (3.17\%) notebooks have \textit{NameError}, which occurs when a declared variable in the notebook is not defined.
374 (8.97\%) notebooks have FileNotFoundError or IOError.
These exceptions occur when absolute paths are used to access data or when the data files are not included in the repository.
\FIG{Figure_timeline_exceptions_per_article} shows how the top three common exceptions \textit{ModuleNotFoundError}, \textit{ImportError}, and \textit{FileNotFoundError} change with the year the article was published .
We see an increase in the \textit{ModuleNotFoundError} through the years.
In the years before 2019, the \textit{ImportError} outnumbered \textit{ModuleNotFoundError}.

\begin{figure}[!htb]
\includegraphics[width=0.8\linewidth]{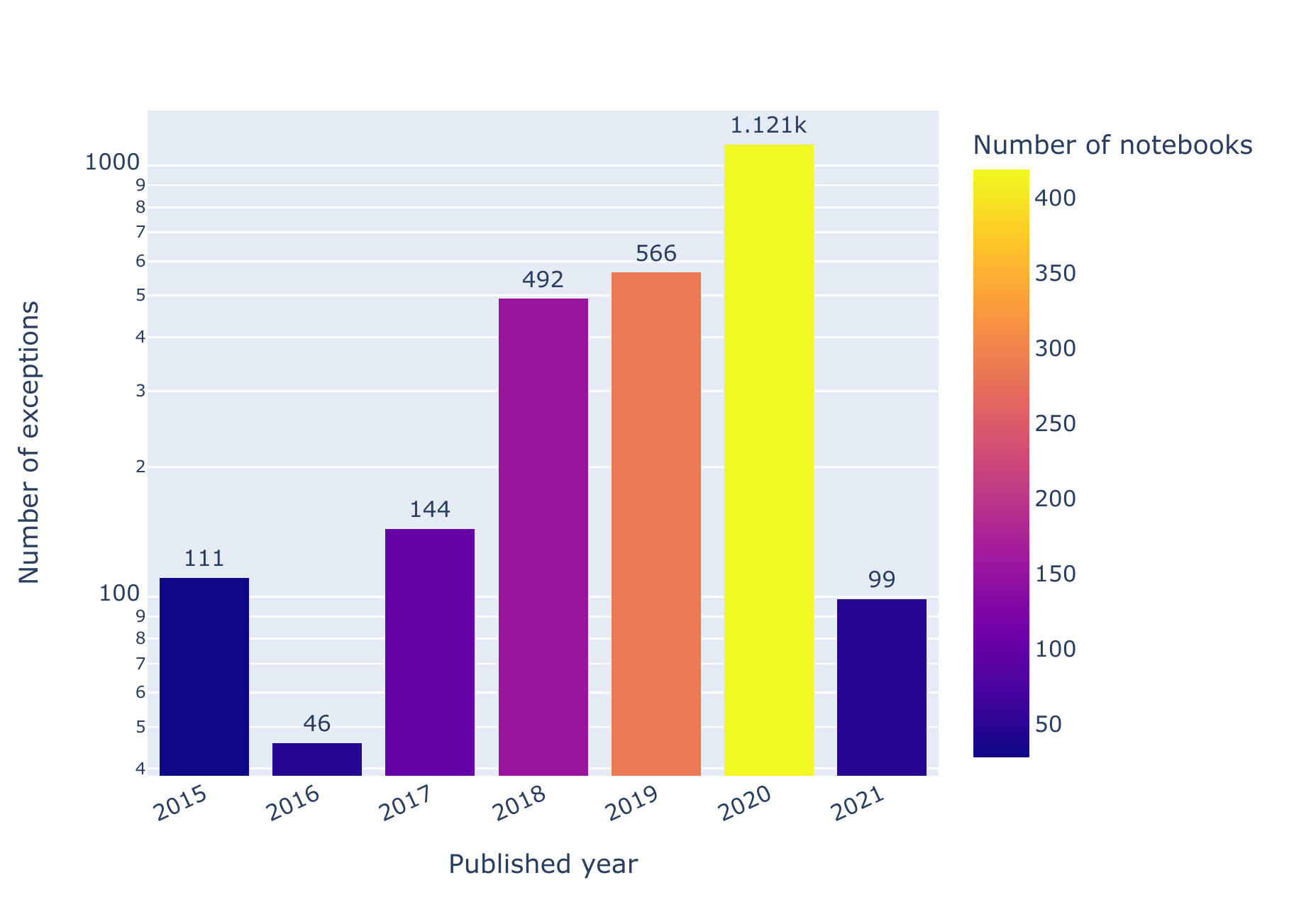}
\caption{Exceptions by year of publication. }
\label{fig:Figure_timeline_exceptions_by_year_notebook}
\end{figure}
\begin{figure}[!htb]
\includegraphics[width=0.8\linewidth]{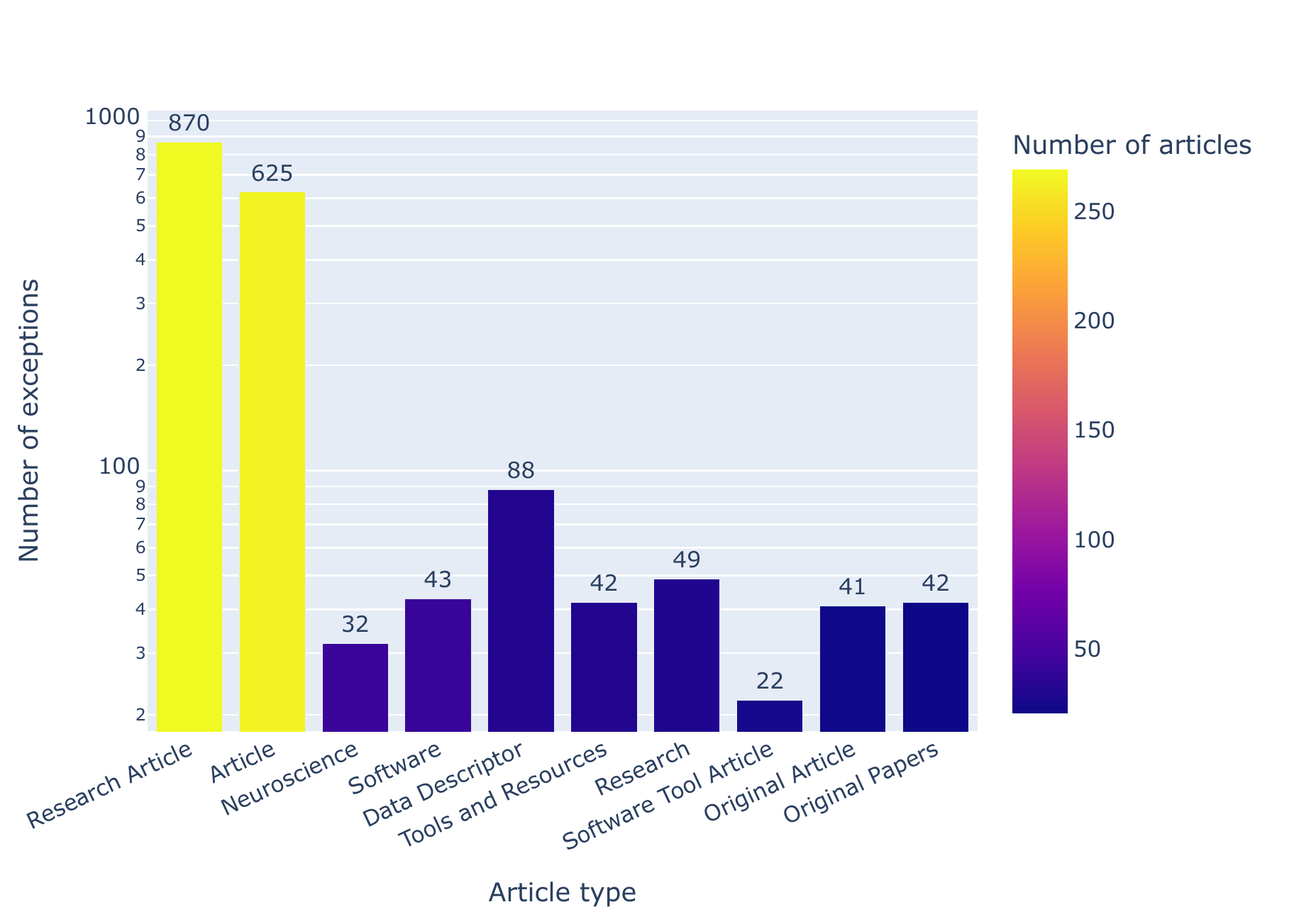}
\caption{Exceptions by article type. }
\label{fig:Figure_exceptions_by_subject_article}
\end{figure}
\begin{figure}[!htb]
\includegraphics[width=0.8\linewidth]{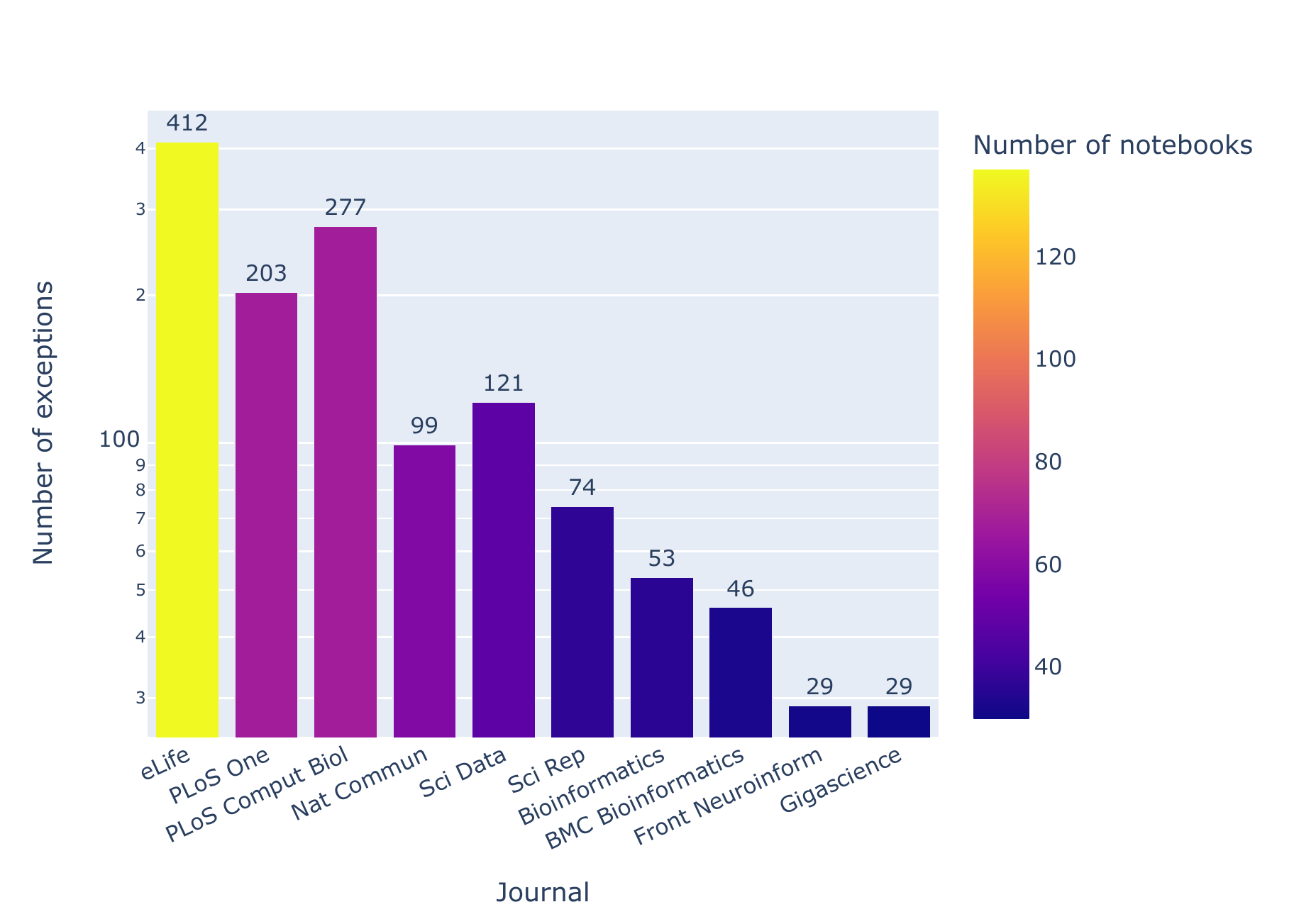}
\caption{Exceptions by journal.}
\label{fig:Figure_exceptions_by_journal_notebooks}
\end{figure}

\FIG{Figure_timeline_exceptions_by_year_notebook} shows the trend of exceptions by the year of publication normalized by the number of notebooks. 
In 2020, we observed the highest number of exceptions and notebooks.
\FIG{Figure_exceptions_by_subject_article} shows the exceptions by the type of the article.
The research articles have the most number of exceptions.
\FIG{Figure_exceptions_by_journal_notebooks} shows the exceptions by journal, normalized by the number of notebooks.
The journal eLife has the most number of notebooks with the most number of exceptions, followed by PLoS One. 

\subsection{Successful replications}
\label{sec:Successful}
396 (9.50\%) of the notebooks in our corpus finished their execution successfully without any errors.
However, for 151 notebooks (3.62\%), our execution generated results that differed from the original ones, while 245 notebooks (5.88\%) produced the same results in our execution as documented for the original notebooks.
\begin{table}[!htb]
\caption[Comparison of notebooks that were successfully executed without errors]{Comparison of notebooks that were successfully executed without errors, grouped by whether their results were \texttt{different} from or \texttt{identical} to the results documented for the original notebook. For features listed in \textit{italics}, the mean values per notebook are indicated, otherwise totals across all notebooks per group.  }
\label{tab:comparison}
\begin{tabular}{| p{0.5\linewidth} | p{0.2\linewidth} | p{0.2\linewidth} |}
\toprule
Features & Notebooks with \texttt{different} results  & Notebooks with \texttt{identical} results \\
\midrule
Number of notebooks     & 151    &  245   \\
\hline
setup.py     & 0    &  98   \\
\hline
requirement.txt     & 0    &  107   \\
\hline
pipfile     & 0    &  0   \\
\hline
\textit{Total cells}     & 17.9    &  17.1   \\
\hline
\textit{Code cells}    & 12.3    &  9.8   \\
\hline
\textit{Markdown cells}    & 5.6    &  6.7   \\
\hline
\textit{Ratio of Markdown vs. code cells}    & 0.46    &  0.68   \\
\hline
\textit{Empty cells}    & 0.7    &  0.7   \\
\hline
\textit{Differences}     & 5.3    &  0   \\
\hline
\textit{Execution time (s)}     & 22.1    &  16.4   \\
\hline
\textit{Execution time per code cell (s)}    & 1.80    &  1.67   \\

\bottomrule
\end{tabular}
\end{table}

Table \ref{tab:comparison} zooms in on the successfully executed notebooks 
and compares those that did not yield results the same results as the original ones (\texttt{different} group) with those that did (\texttt{identical} group). 
A clear difference between both groups is that many of the notebooks in the \texttt{identical} group had their dependencies specified via either setup.py or requirements.txt or both, in contrast to none of the notebooks in the \texttt{different} group. Since notebooks with no dependency declarations were run using the default conda dependencies, the fact that they successfully finished means that all dependencies were covered. However, as the version of the dependencies used in the original notebook was not documented, it may have differed from the version that was provided in our respective Conda environment. 

Besides versioning of dependencies, there could be a number of other reasons as to why an error-free execution might yield different results. For instance, random functions may be invoked, or code cells in the original might have been executed multiple times or in a different order than in our execution, which ran every code cell just once, from top to bottom.
However, we would not expect the invocation of random functions or an inconsistent execution order to correlate so strongly with whether the dependencies had been explicitly declared or not. 

In contrast to the dependency declarations, other features in Table \ref{tab:comparison} show more gradual differences between the two groups, and they largely fit with intuition. For instance, it is understandable that notebooks with more code cells take longer to execute and that code whose execution per code cell takes longer is somewhat more complex, thus raising the probability of different outcomes. It is also not surprising that, while the total number of cells per notebook is nearly the same in both groups, notebooks in the \texttt{identical} group show a higher ratio of Markdown versus code cells, since that ratio is indicative of documentation efforts, and better documentation would be expected to go with better reproducibility.

The average number of differences observed per notebook (or even per code cell) is not easy to interpret on its own, as it includes differences in output cells, cell counter values or in output files, and a difference early in a notebook can lead to further differences later.

Table \ref{tab:versioncomparison} illustrates how different Python versions performed in terms of successful executions: amongst the top 5 versions for notebooks yielding different results, there were three 2.7 versions, whereas there were three 3.7 versions in the group that yielded identical results, and 3.6.9 and 3.6.5 were represented roughly equally in both groups. 

\begin{table}[!htb]
\caption[Comparison of most frequent Python versions declared for notebooks that were successfully executed without errors.]{Comparison of most frequent Python versions declared for notebooks that were successfully executed without errors, grouped by whether their results were \texttt{different} from or \texttt{identical} to the results documented for the original notebook. Versions listed in \textit{italics} occur in both top-5 groups, versions listed in \textbf{bold} in only one. The \texttt{absolute} columns give total number of notebooks per version and group, while the \texttt{relative} columns normalize the absolute values as a percentage of the total number of notebooks per group, i.e. 151 for \texttt{different} and 245 for \texttt{identical}, as per Table \ref{tab:comparison}. In both groups, the top-5 versions account for slightly over half of the notebooks. }
\label{tab:versioncomparison}
\begin{tabular}{|r|lrr|lrr|}
\toprule
\multicolumn{1}{|l|}{}     & \multicolumn{3}{c|}{\texttt{different}}                                                                       & \multicolumn{3}{c|}{\texttt{identical}}                                                                      \\ \hline\multicolumn{1}{|r|}{rank} & \multicolumn{1}{r|}{version}         & \multicolumn{1}{l|}{absolute} & \multicolumn{1}{r|}{relative} & \multicolumn{1}{r|}{version}        & \multicolumn{1}{r|}{absolute} & \multicolumn{1}{r|}{relative} \\ \midrule
1                          & \multicolumn{1}{l|}{\textit{3.6.9}}  & \multicolumn{1}{r|}{28}       & 18.5                          & \multicolumn{1}{r|}{\textbf{3.7.3}} & \multicolumn{1}{r|}{46}       & 18.8                          \\ \hline
2                          & \multicolumn{1}{l|}{\textbf{2.7.6}}  & \multicolumn{1}{r|}{17}       & 11.3                          & \multicolumn{1}{r|}{\textit{3.6.9}} & \multicolumn{1}{r|}{30}       & 12.2                          \\ \hline
3                          & \multicolumn{1}{l|}{\textbf{2.7.10}} & \multicolumn{1}{r|}{13}       & 8.6                           & \multicolumn{1}{r|}{\textbf{3.7.1}} & \multicolumn{1}{r|}{26}       & 10.6                          \\ \hline
4                          & \multicolumn{1}{l|}{\textit{3.6.5}}  & \multicolumn{1}{r|}{12}       & 7.9                           & \multicolumn{1}{r|}{\textbf{3.7.4}} & \multicolumn{1}{r|}{22}       & 9.0                           \\ \hline
5                          & \multicolumn{1}{l|}{\textbf{2.7.9}}  & \multicolumn{1}{r|}{11}       & 7.3                           & \multicolumn{1}{r|}{\textit{3.6.5}} & \multicolumn{1}{r|}{15}       & 6.1    \\ \bottomrule
\end{tabular}
\end{table}

Other parameters that we considered but did not include in the analysis of the finished notebooks were the number of dependencies (the more there are, the more likely replicability is reduced; see also section
\nameref{sec:NotebookDependencies} and in particular
Fig.\ \ref{fig:Figure_dependencies}), the type of dependencies (e.g. local code or environment,  Python package, local or remote file or service, each of which could complicate replication; see also 
Fig.\ 
\ref{fig:Figure_f_notebook_module_full_import}
and
\ref{fig:Figure_f_notebook_module_load_ext_full_import}), the recency (cf.\ Fig.\ and \ref{fig:Figure_timeline_exceptions_per_article}
and
\ref{fig:Figure_timeline_exceptions_by_year_notebook}) 
of the notebooks (more recent ones would be expected to be more replicable)
or notebook titles 
(cf.\ Fig.\ \ref{fig:Figure_notebook_name_count}, \ref{fig:Figure_f_notebook_tests} and
\ref{fig:Figure_f_notebook_name_length_full})
containing strings like  ``tutorial'' or ``demo'' (which might be indicative of expected reuse, thus perhaps triggering more careful documentation) or ``untitled'' (which is the default title and may thus indicate a lack of attention to documentation and, consequently, a higher likelihood for replication attempts to fail).

\subsection{Notebook Styling}
\label{sec:NotebookStyling}

\begin{figure}[!htb]
\includegraphics[width=0.8\linewidth]{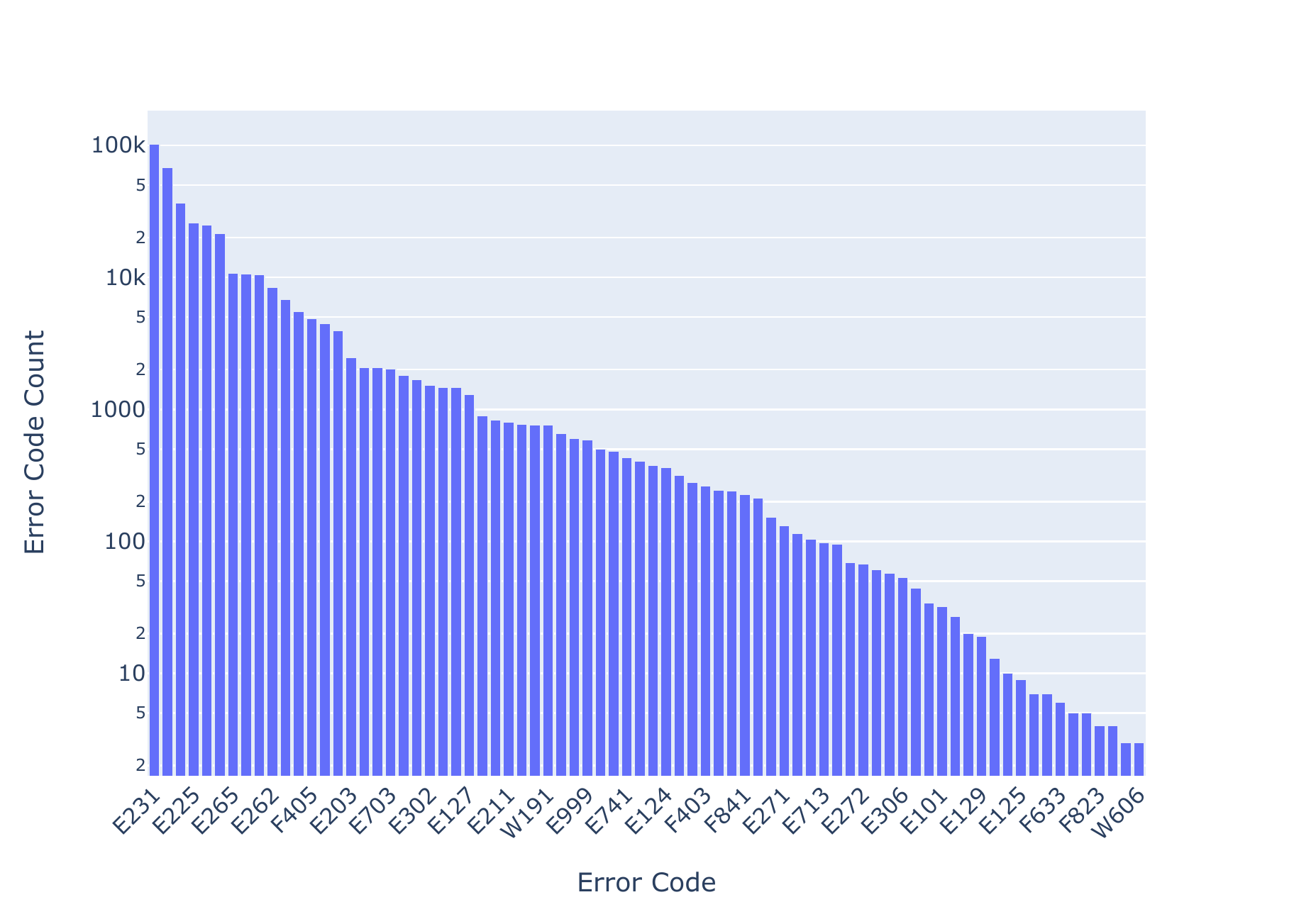}
\caption{Frequent notebook code style errors as per the Python code style guide. 
}
\label{fig:Figure_top10_notebook_codestyle_exceptions}
\end{figure}

In addition to the common exceptions in the notebooks, we also checked the notebook code styling errors.
\FIG{Figure_top10_notebook_codestyle_exceptions} shows the most common Python code warning/style errors found in our study.

\begin{table}[!htb]
\caption{Common Python Notebook Code Warning/Style Error found in our Study }
\label{tab:pycodestyling}
\begin{tabular}{| p{0.1\linewidth} | p{0.5\linewidth} | p{0.2\linewidth} |}
\toprule
Error code & Description & Count\\
\midrule
E231     & missing whitespace after commas, semicolons or colons & 102218  \\
E225     & missing whitespace around operator  &  25979 \\
E265     & block comment should start with ‘\#‘  & 10769 \\
E402    & module level import not at top of file & 10478 \\
E262     & inline comment should start with ‘\#‘ & 8369 \\
E703     & statement ends with a semicolon  & 2023 \\
E127    & continuation line over-indented for visual indent & 1290 \\
E701 	& multiple statements on one line & 500 \\
E741    & do not use variables named ‘l’, ‘O’, or ‘I’ & 432 \\
E401 	& multiple imports on one line & 95 \\
E101 	& indentation contains mixed spaces and tabs & 32 \\
\hline
F405     & \textit{name} may be undefined, or defined from star imports: \textit{module}  & 4840 \\
F401    & \textit{module} imported but unused & 3938 \\
F821    & undefined name 'X'  & 2071 \\
F403    & 'from module import *' used; unable to detect undefined names & 263 \\
F841    & local variable 'X' is assigned to but never used & 225 \\
F404    & future import(s) name after other statements & 44 \\
F402    & import 'X' from line Y shadowed by loop variable & 10 \\ 
F633    & use of >> is invalid with print function & 6 \\
F823    & local variable 'X' defined in enclosing scope on line Y referenced before assignment & 4 \\
\hline
W601 	& .has\_key() is deprecated, use ‘in’ & 7 \\
W606    & 'async' and 'await' are reserved keywords starting with Python 3.7 & 3 \\
\bottomrule
\end{tabular}
\end{table}
Table \ref{tab:pycodestyling} presents the code for the Python code warnings and style errors found in our study.
E231 is the most common coding style error, followed by E225 and E265, respectively.
There are also some common content errors other than styling errors like F403 and F405~-- these are related to variable and module definition errors.
The W601 and W606 warnings relate to the use of deprecated and reserved keys.

\newpage
\section{Discussion}
\label{sec:Discussion}

In this study, we have analyzed 
the \textit{Method reproducibility}~--~in the sense of \citet{goodman2016what}~--~
of Jupyter notebooks written in Python and publicly hosted on GitHub that are mentioned in publications whose full text was available via PubMed Central by the reference period, i.e.\ the time when our reproducibility pipeline was run on 24-28 February 2021. We will now discuss the limitations of the study and then its implications, again primarily for \textit{Method reproducibility} of Jupyter notebooks associated with biomedical publications.

\subsection{Limitations}
\label{subsec:Limitations}

The present study does not address \textit{Inferential reproducibility} and only briefly touches upon  \textit{Results reproducibility}. Furthermore, we made no attempt to re-run
computational notebooks that met any of the following exclusion criteria during the reference period: (a) they did not use Jupyter (or its precursor, IPython), (b) they were not written in Python, (c) they were not publicly available on GitHub, (d) they were not mentioned in publications available from PubMed Central, (e) they were not on the base branch of their GitHub repository (which is the only branch we looked at).

Our reproducibility workflow is based on that by \cite{pimentel2019a},
with some changes to include GitHub repositories from publications and using the \texttt{nbdime} library 
\citep{nbdime}
from Jupyter instead of string matching for finding differences in the notebook outputs.
The approach is using conda \citep{Conda}
environments. We did not use any Docker images \citep{Docker2013}
for the execution environment, even in cases when they were available.
This workflow being fully automated, we did not spend any manual effort on fixing any of the errors that came up for an individual notebook~-- see \citet{woodbridge2017jupyter} 
for a report of an attempt to do so, which also provided the foundation for a prototypical validation tool that makes use of GitLab Actions 

For a good number of the reported problems (especially the missing software or data dependencies, as per Fig.\ \ref{fig:Figure_top_exception_by_reason}),
it is often straightforward to fix them manually for individual notebooks, yet undertaking manual fixes systematically was not practical at the scale of the thousands of notebooks rerun here. 
If the original code had specified dependencies without referring to a specific version, our rerun would use the most recent 
conda-installable
version of that library.
Finally, in estimating the environmental footprint of this study, we only included the footprint due to running the full pipeline once~-- we did not include the efforts involved in preparing the pipeline, analyzing the data or writing the manuscript.

\subsection{Implications}
There are several implications of this study. 
First, on a general level, the low degree of reproducibility that we documented here for Jupyter notebooks associated with biomedical publications 
goes conform with similarly low levels of reproducibility that were found in earlier domain-generic studies, both for Python \citep{rule2018exploration,pimentel2021understanding}
and R
\citep{trisovic2022large}.

Second, considering that the notebooks we explored here were associated with peer-reviewed publications, it is clear that the review processes currently in place at journals within our corpus does not generally pay much attention to the reproducibility of the notebooks. 
This clearly needs to change, and we need systemic approaches to that rather than just adding this to the list of things the reviewers are expected to attend to. As our study demonstrates, a basic level of reproducibility assessment can well be achieved in a fully automated fashion, so it would probably be beneficial in terms of research quality to include such automated basic checks~-- for notebooks and other software~-- into standard review procedures. Ideally, this would be done in a way that works across publishers as well as for a variety of technology stacks and programming languages.

Third, 
while there is a large variety in the types of errors affecting reproducibility, some of the most common errors concentrate around dependencies (cf.\ Fig.\  \ref{fig:Figure_f_notebook_module_full_import},
\ref{fig:Figure_f_notebook_module_load_ext_full_import}, \ref{fig:Figure_dependencies} and \ref{fig:Figure_top_exception_by_reason}), so efforts aimed at systemic improvements of dependency handling~-- e.g.\ as per \cite{zhu2021restoring}~--
have the potential to increase reproducibility considerably. 
Here, programming language-specific efforts regarding code dependencies can be combined with efforts targeted at improving the automated handling of data dependencies, which would be beneficial irrespective of the specific programming language.

Fourth, zooming in on Python specifically, wider adoption of existing workflows for code dependency management (such as \texttt{requirements.txt}) as well as associated checks during the publishing process would help. Researchers attempting to publish research with associated notebooks should not have to do this all by themselves~-- research infrastructures as well as publishers and funders can all help facilitate establishing best practice here and engaging communities around them.

Fifth, the few notebooks that actually did reproduce (cf.\  \nameref{sec:Successful})
are not equally distributed.
This means that reproducibility could probably be strengthened by enhancing or highlighting  the features that correlate with it. For instance, Jupyter notebooks with more emphasis on documentation scored better than others, and there is merit in the idea of making Jupyter notebooks or similar computational notebooks a publication type of their own. This is already the case in some places, as examplified by \citet{constantine2016python} or \cite{garg2022pygetpapers} in the Journal of Open Source Software.

Sixth, the ongoing diversification of the Jupyter ecosystem~-- e.g.\ in terms of programming languages, deployment frameworks or cloud infrastructure~--
is increasingly reflected, albeit with delay, in the biomedical literature. In parallel, while GitHub remains hugely popular, alternatives like GitLab, Gitee or Codeberg are growing too. 
Future assessments of Jupyter reproducibility will thus need to take this increasing complexity into account, and ideally present some systematic approach to it.

Seventh, the delays that come with current publishing practices also mean that Jupyter notebooks associated with freshly published papers are using software versions near or even beyond their respective support window (which is 42 months in much of the Python ecosystem\footnote{Cf.\ \href{https://numpy.org/neps/nep-0029-deprecation_policy.html}{https://numpy.org/neps/nep-0029-deprecation\_policy.html}}).

For instance, the oldest Python version still officially supported in 2021 was 3.6 (which was itself retired by the end of 2021, when 3.10 was released\footnote{See \href{https://endoflife.date/python}{https://endoflife.date/python} for release schedule}), yet as shown in Figure
\ref{fig:Figure_timeline_python_minor_version_by_repo_update}, 
over a thousand Python notebooks in our corpus whose last commit was in 2021 still featured earlier Python versions, mainly 2.7 (outphased in 2020) but also 3.4 (2019), 3.5 (2020) and some for which the version could not be determined. This contributes to reproducibility issues. A similar issue exists with the versions of the libraries called from any given notebook, though the effects might differ as a function of whether they have been invoked with or without the version being specified. If the version had been specified, its official end of life might go back even further. If the version was not specified, the newest available version would be invoked, which may not be compatible with the way the library had been used in the original notebook. 
Similar issues can arise with the versioning of APIs, datasets, ontologies or other standards used in the notebook, all of which can contribute to reduced reproducibility.
To some extent, these version delay issues can be shortened by preprints: since they are (essentially by definition, but not always in practice) published before the final version of the associated manuscript, and hence their delays should be shorter, with lower reductions in reproducibility, though we did not investigate that in detail. 

Eight, the variety and scale of issues encountered in the notebooks analyzed here provides ample opportunities for use in educational contexts~-- including instructed, self-guided or group learning~-- since fixing real-life bugs can be more motivating than working primarily with textbook examples.
To do this effectively would require some mapping of the strengths and weaknesses of the notebooks to learning objectives, which may range from understanding programming paradigms, software engineering principles or data integration workflows to developing an appreciation for documentation and other aspects of good scientific practice.
Given the continuously expanding breadth of publications that use Jupyter notebooks, it is also steadily becoming easier to find publications where they have been used in research meeting specific criteria. These could be a particular topic~-- 
e.g.\
natural products research \citep{mayr2020finding}
or invasion biology \citep{bors2019population}~-- or workflows involving a particular experimental
methodology
like single-cell RNA sequencing \citep{vargo2020rank} or other software tools like ImageJ \citep{bryson2020composite}.

\section{Conclusions}
\label{sec:Conclusions}

On the basis of re-running 
4169
Jupyter notebooks associated with 
1419
publications whose full text is available via Pubmed Central, we conclude that
such notebooks are becoming more and more popular for sharing code associated with biomedical publications, that the range of programming languages or journals they cover is continuously expanding
and that their reproducibility is low but improving,
consistent with earlier studies on Jupyter notebooks shared in other contexts.
The main issues are related to dependencies~-- both code and data~-- which means that reproducibility could likely be improved considerably if the code~-- and dependencies in particular~-- were better documented. Further improvements could be expected if some basic and automated reproducibility checks of the kind performed here
were to be systematically included in the peer review process or if computational notebooks~-- Jupyter or otherwise~-- were combined  
 with additional approaches that address reproducibility from other angles, e.g. registered reports.

\section{Data availability}
\label{sec:Dataavailability}
All the data generated during this study is available at \url{https://doi.org/10.5281/zenodo.6802158}.
The code used is available at \url{https://github.com/fusion-jena/computational-reproducibility-pmc}.
The code contains notebooks used for analysis of the results.

\section{Conflict of interest}
The authors declare there are no competing interests.

\section{Ethics}
No facet of the research reported here triggered a requirement for ethical review. While our data contains personally identifiable information, it was taken directly from PMC.
We did, however, consider the ethical implications of automated reproducibility studies of the kind presented here, which led us to (a) highlight systemic aspects, (b) not zoom in on individual stakeholders other than journals and (c) include environmental footprint information.

\section{Acknowledgements}
We would like to thank the providers of infrastructure, data and code that we used in this study. These include the PubMed Central repository hosted by the National Center for Biotechnology Information in the United States and the Ara Cluster at the University of Jena as well as the Python, Jupyter and Conda communities and their respective dependencies.
We acknowledge the 
Open Research Doathon
on the occasion of Open Data Day 2017, where the first attempts at systematic reproduction of PMC-indexed Jupyter notebooks were made \cite{woodbridge2017jupyter}.
Special thanks go to JupyterCon, which made the two of us aware of each other's work and provided the nucleus for our collaboration.
Work by S.S. was supported 
by the Carl Zeiss Foundation for the project ``A Virtual Werkstatt for Digitization in the Sciences (K3)'' \citep{samuel2020virtual} within the scope of the program line ``Breakthroughs: Exploring Intelligent Systems for Digitization - explore the basics, use applications''. Work by D.M. was supported by the Alfred P. Sloan Foundation under grant number G-2021-17106.
The computational experiments were performed on resources of Friedrich Schiller University Jena supported in part by DFG grants INST 275/334-1 FUGG and INST 275/363-1 FUGG.

\bibliography{jupyter-reproducibility.bib}

\begin{thebibliography}{79}
\providecommand{\natexlab}[1]{#1}
\providecommand{\urlprefix}{}
\providecommand{\doiprefix}{doi: }

\bibitem[{Bairoch(2018)Bairoch, Amos Marc}]{bairoch2018cellosaurus}
\textbf{\color{eLifeMediumGrey} Bairoch AM}.
\newblock The Cellosaurus, a Cell-Line Knowledge Resource.
\newblock Journal of Biomolecular Techniques.  2018; 29(2):25--38.

\bibitem[{Baker et~al.(2019)Baker, Ann-Marie and Cereser, Biancastella and
  Melton, Samuel and Fletcher, Alexander G. and Rodriguez-Justo, Manuel and
  Tadrous, Paul J. and Humphries, Adam and Elia, George and McDonald, Stuart
  A.C. and Wright, Nicholas A. and Simons, Benjamin D. and Jansen, Marnix and
  Graham, Trevor A.}]{baker2019quantification}
\textbf{\color{eLifeMediumGrey} Baker AM}, Cereser B, Melton S, Fletcher AG,
  Rodriguez-Justo M, Tadrous PJ, Humphries A, Elia G, McDonald SAC, Wright NA,
  Simons BD, Jansen M, Graham TA.
\newblock Quantification of {Crypt} and {Stem} {Cell} {Evolution} in the
  {Normal} and {Neoplastic} {Human} {Colon}.
\newblock Cell Reports.  2019 may 1; 27(8):2524.

\bibitem[{Baker(2016)Baker, Monya}]{baker20161500}
\textbf{\color{eLifeMediumGrey} Baker M}.
\newblock 1,500 scientists lift the lid on reproducibility.
\newblock Nature.  2016 may 25; 533(7604):452--454.

\bibitem[{Boettiger(2015)Boettiger, Carl}]{boettiger2015Docker}
\textbf{\color{eLifeMediumGrey} Boettiger C}.
\newblock An Introduction to Docker for Reproducible Research.
\newblock SIGOPS Oper Syst Rev.  2015 Jan; 49(1):71--79.
\newblock \urlprefix\url{http://doi.acm.org/10.1145/2723872.2723882},
  \href{10.1145/2723872.2723882}{\doiprefix
  \detokenize{10.1145/2723872.2723882}}.

\bibitem[{Bors et~al.(2019)Bors, Eleanor K and Herrera, Santiago and Morris,
  James A and Shank, Timothy M}]{bors2019population}
\textbf{\color{eLifeMediumGrey} Bors EK}, Herrera S, Morris JA, Shank TM.
\newblock Population genomics of rapidly invading lionfish in the {Caribbean}
  reveals signals of range expansion in the absence of spatial population
  structure.
\newblock Ecology and evolution.  2019 feb 10; 9(6):3306--3320.

\bibitem[{Brito et~al.(2020)Brito, Jaqueline J and Li, Jun and Moore, Jason H
  and Greene, Casey S and Nogoy, Nicole A and Garmire, Lana X and Mangul,
  Serghei}]{brito2020recommendations}
\textbf{\color{eLifeMediumGrey} Brito JJ}, Li J, Moore JH, Greene CS, Nogoy NA,
  Garmire LX, Mangul S.
\newblock {Recommendations to enhance rigor and reproducibility in biomedical
  research}.
\newblock GigaScience.  2020 06; 9(6).
\newblock \urlprefix\url{https://doi.org/10.1093/gigascience/giaa056},
  \href{10.1093/gigascience/giaa056}{\doiprefix
  \detokenize{10.1093/gigascience/giaa056}}, giaa056.

\bibitem[{Bryson et~al.(2020)Bryson, Abigail E and Brown, Maya Wilson and
  Mullins, Joey and Dong, Wei and Bahmani, Keivan and Bornowski, Nolan and
  Chiu, Christina and Engelgau, Philip and Gettings, Bethany and Gomezcano,
  Fabio and Gregory, Luke M and Haber, Anna C and Hoh, Donghee and Jennings,
  Emily E and Ji, Zhongjie and Kaur, Prabhjot and Raju, Sunil K Kenchanmane and
  Long, Yunfei and Lotreck, Serena G and Mathieu, Davis T and Ranaweera,
  Thilanka and Ritter, Eleanore J and Sadohara, Rie and Shrote, Robert Z and
  Smith, Kaila E and Teresi, Scott J and Venegas, Julian and Wang, Hao and
  Wilson, McKena L and Tarrant, Alyssa R and Frank, Margaret H and Migicovsky,
  Zo{\" e} and Kumar, Jyothi and VanBuren, Robert and Londo, Jason P and
  Chitwood, Daniel H}]{bryson2020composite}
\textbf{\color{eLifeMediumGrey} Bryson AE}, Brown MW, Mullins J, Dong W,
  Bahmani K, Bornowski N, Chiu C, Engelgau P, Gettings B, Gomezcano F, Gregory
  LM, Haber AC, Hoh D, Jennings EE, Ji Z, Kaur P, Raju SKK, Long Y, Lotreck SG,
  Mathieu DT, et~al.
\newblock Composite modeling of leaf shape along shoots discriminates {Vitis}
  species better than individual leaves.
\newblock Applications in plant sciences.  2020 dec 3; 8(12):e11404.

\bibitem[{Burlingame et~al.(2021)Burlingame, Erik A. and Eng, Jennifer and
  Thibault, Guillaume and Chin, Koei and Gray, Joe W. and Chang, Young
  Hwan}]{burlingame2021toward}
\textbf{\color{eLifeMediumGrey} Burlingame EA}, Eng J, Thibault G, Chin K, Gray
  JW, Chang YH.
\newblock Toward reproducible, scalable, and robust data analysis across
  multiplex tissue imaging platforms.
\newblock Cell Reports Methods.  2021 8; 1(4):100053.

\bibitem[{Chattopadhyay et~al.(2020)Chattopadhyay, Souti and Prasad, Ishita and
  Henley, Austin Z and Sarma, Anita and Barik, Titus}]{chattopadhyay2020s}
\textbf{\color{eLifeMediumGrey} Chattopadhyay S}, Prasad I, Henley AZ, Sarma A,
  Barik T.
\newblock What's Wrong with Computational Notebooks? Pain Points, Needs, and
  Design Opportunities.
\newblock In: \emph{Proceedings of the 2020 CHI Conference on Human Factors in
  Computing Systems}; 2020. p. 1--12.

\bibitem[{Chirigati et~al.(2013)Fernando Chirigati and Dennis Shasha and
  Juliana Freire}]{chirigati2013reprozip}
\textbf{\color{eLifeMediumGrey} Chirigati F}, Shasha D, Freire J.
\newblock Repro{Z}ip: Using Provenance to Support Computational
  Reproducibility.
\newblock In: \emph{Presented as part of the 5th {USENIX} Workshop on the
  Theory and Practice of Provenance} Lombard, IL: {USENIX}; 2013.
  \urlprefix\url{https://www.usenix.org/conference/tapp13/reprozip-using-provenance-support-computational-reproducibilitythe}.

\bibitem[{Cock et~al.(2009)Cock, Peter J. A. and Antao, Tiago and Chang,
  Jeffrey T. and Chapman, Brad A. and Cox, Cymon J. and Dalke, Andrew and
  Friedberg, Iddo and Hamelryck, Thomas and Kauff, Frank and Wilczynski, Bartek
  and de Hoon, Michiel J. L.}]{cock2009biopython}
\textbf{\color{eLifeMediumGrey} Cock PJA}, Antao T, Chang JT, Chapman BA, Cox
  CJ, Dalke A, Friedberg I, Hamelryck T, Kauff F, Wilczynski B, de~Hoon MJL.
\newblock {Biopython: freely available Python tools for computational molecular
  biology and bioinformatics}.
\newblock Bioinformatics.  2009 03; 25(11):1422--1423.
\newblock \urlprefix\url{https://doi.org/10.1093/bioinformatics/btp163},
  \href{10.1093/bioinformatics/btp163}{\doiprefix
  \detokenize{10.1093/bioinformatics/btp163}}.

\bibitem[{Coiera et~al.(2018)Coiera, Enrico W. and Ammenwerth, Elske and
  Georgiou, Andrew and Magrabi, Farah}]{coiera2018does}
\textbf{\color{eLifeMediumGrey} Coiera EW}, Ammenwerth E, Georgiou A, Magrabi
  F.
\newblock Does health informatics have a replication crisis?
\newblock Journal of the American Medical Informatics Association.  2018 aug 1;
  25(8):963--968.
\newblock \href{10.1093/JAMIA/OCY028}{\doiprefix
  \detokenize{10.1093/JAMIA/OCY028}}.

\bibitem[{{Conda community}(2017)}]{Conda}
\textbf{\color{eLifeMediumGrey} {Conda community}}, Conda; 2017.
\newblock \urlprefix\url{https://conda.io/}.

\bibitem[{Constantine et~al.(2016)Constantine, Paul and Howard, Ryan and Glaws,
  Andrew and Grey, Zachary and Diaz, Paul and Fletcher,
  Leslie}]{constantine2016python}
\textbf{\color{eLifeMediumGrey} Constantine P}, Howard R, Glaws A, Grey Z, Diaz
  P, Fletcher L.
\newblock Python {Active}-subspaces {Utility} {Library}.
\newblock Journal of Open Source Software.  2016 sep 29; 1(5):79.

\bibitem[{Contera(2021)Contera, Sonia}]{contera2021communication}
\textbf{\color{eLifeMediumGrey} Contera S}.
\newblock Communication is central to the mission of science.
\newblock Nature Reviews Materials.  2021; 6(5):377--378.

\bibitem[{Crick et~al.(2017)Crick, Tom and Hall, Benjamin A. and Ishtiaq,
  Samin}]{crick2017reproducibility}
\textbf{\color{eLifeMediumGrey} Crick T}, Hall BA, Ishtiaq S.
\newblock Reproducibility in {Research}: Systems, {Infrastructure}, {Culture}.
\newblock Journal of open research software.  2017 nov 9; 5(1):32.

\bibitem[{Docker(2013)}]{Docker2013}
\textbf{\color{eLifeMediumGrey} Docker}, Docker; 2013.
\newblock \urlprefix\url{https://www.docker.com}.

\bibitem[{Fanelli(2018)Fanelli, Daniele}]{fanelli2018opinion}
\textbf{\color{eLifeMediumGrey} Fanelli D}.
\newblock Opinion: Is Science Really Facing a Reproducibility Crisis, and Do We
  Need It To?
\newblock Proceedings of the National Academy of Sciences of the United States
  of America.  2018; 115(11):2628--2631.

\bibitem[{Garg et~al.(2022)Garg, Ayush and Smith-Unna, Richard D and
  Murray-Rust, Peter}]{garg2022pygetpapers}
\textbf{\color{eLifeMediumGrey} Garg A}, Smith-Unna RD, Murray-Rust P.
\newblock pygetpapers: a {Python} library for automated retrieval of scientific
  literature.
\newblock Journal of Open Source Software.  2022 jul 7; 7(75):4451.

\bibitem[{Gil et~al.(2016)Gil, Yolanda and David, Cédric H. and Demir, Ibrahim
  and Essawy, Bakinam T. and Fulweiler, Robinson W. and Goodall, Jonathan L.
  and Karlstrom, Leif and Lee, Huikyo and Mills, Heath J. and Oh, Ji-Hyun and
  Pierce, Suzanne A. and Pope, Allen and Tzeng, Mimi W. and Villamizar, Sandra
  R. and Yu, Xuan}]{gil2016toward}
\textbf{\color{eLifeMediumGrey} Gil Y}, David CH, Demir I, Essawy BT, Fulweiler
  RW, Goodall JL, Karlstrom L, Lee H, Mills HJ, Oh JH, Pierce SA, Pope A, Tzeng
  MW, Villamizar SR, Yu X.
\newblock Toward the Geoscience Paper of the Future: Best practices for
  documenting and sharing research from data to software to provenance.
\newblock Earth and Space Science.  2016; 3(10):388--415.
\newblock
  \urlprefix\url{https://agupubs.onlinelibrary.wiley.com/doi/abs/10.1002/2015EA000136},
  \href{10.1002/2015EA000136}{\doiprefix \detokenize{10.1002/2015EA000136}}.

\bibitem[{Goodman et~al.(2016)Goodman, Steven N and Fanelli, Daniele and
  Ioannidis, John P A}]{goodman2016what}
\textbf{\color{eLifeMediumGrey} Goodman SN}, Fanelli D, Ioannidis JPA.
\newblock What does research reproducibility mean?
\newblock Science Translational Medicine.  2016 jun 1; 8(341):341ps12.
\newblock \href{10.1126/scitranslmed.aaf5027}{\doiprefix
  \detokenize{10.1126/scitranslmed.aaf5027}}.

\bibitem[{Granger and Perez(2021)Granger, Brian E. and Perez,
  Fernando}]{granger2021jupyter}
\textbf{\color{eLifeMediumGrey} Granger BE}, Perez F.
\newblock Jupyter: Thinking and {Storytelling} {With} {Code} and {Data}.
\newblock Computing in Science and Engineering.  2021 mar 26; 23(2):7--14.

\bibitem[{Gray et~al.(2012)Gray, Steven and Shwom, Rachael and Jordan,
  Rebecca}]{gray2012understanding}
\textbf{\color{eLifeMediumGrey} Gray S}, Shwom R, Jordan R.
\newblock Understanding factors that influence stakeholder trust of natural
  resource science and institutions.
\newblock Environmental management.  2012; 49(3):663--674.

\bibitem[{Gr{\" u}ning et~al.(2018)Gr{\" u}ning, Bj{\" o}rn and Chilton, John
  and K{\" o}ster, Johannes and Dale, Ryan and Soranzo, Nicola and van den
  Beek, Marius and Goecks, Jeremy and Backofen, Rolf and Nekrutenko, Anton and
  Taylor, James}]{gruning2018practical}
\textbf{\color{eLifeMediumGrey} Gr{\" u}ning B}, Chilton J, K{\" o}ster J, Dale
  R, Soranzo N, van~den Beek M, Goecks J, Backofen R, Nekrutenko A, Taylor J.
\newblock Practical {Computational} {Reproducibility} in the {Life} {Sciences}.
\newblock Cell systems.  2018 jun 1; 6(6):631--635.
\newblock \href{10.1016/j.cels.2018.03.014}{\doiprefix
  \detokenize{10.1016/j.cels.2018.03.014}}.

\bibitem[{Guttinger(2020)Guttinger, Stephan}]{guttinger2020limits}
\textbf{\color{eLifeMediumGrey} Guttinger S}.
\newblock The limits of replicability.
\newblock European journal for philosophy of science.  2020 jan 15; 10(2).
\newblock \href{10.1007/s13194-019-0269-1}{\doiprefix
  \detokenize{10.1007/s13194-019-0269-1}}.

\bibitem[{Halchenko et~al.(2021)Yaroslav O. Halchenko and Kyle Meyer and
  Benjamin Poldrack and Debanjum Singh Solanky and Adina S. Wagner and Jason
  Gors and Dave MacFarlane and Dorian Pustina and Vanessa Sochat and Satrajit
  S. Ghosh and Christian Mönch and Christopher J. Markiewicz and Laura Waite
  and Ilya Shlyakhter and Alejandro de la Vega and Soichi Hayashi and Christian
  Olaf Häusler and Jean-Baptiste Poline and Tobias Kadelka and Kusti Skytén
  and Dorota Jarecka and David Kennedy and Ted Strauss and Matt Cieslak and
  Peter Vavra and Horea-Ioan Ioanas and Robin Schneider and Mika Pflüger and
  James V. Haxby and Simon B. Eickhoff and Michael
  Hanke}]{halchenko2021datalad}
\textbf{\color{eLifeMediumGrey} Halchenko YO}, Meyer K, Poldrack B, Solanky DS,
  Wagner AS, Gors J, MacFarlane D, Pustina D, Sochat V, Ghosh SS, Mönch C,
  Markiewicz CJ, Waite L, Shlyakhter I, de~la Vega A, Hayashi S, Häusler CO,
  Poline JB, Kadelka T, Skytén K, et~al.
\newblock DataLad: distributed system for joint management of code, data, and
  their relationship.
\newblock Journal of Open Source Software.  2021; 6(63):3262.
\newblock \urlprefix\url{https://doi.org/10.21105/joss.03262},
  \href{10.21105/joss.03262}{\doiprefix \detokenize{10.21105/joss.03262}}.

\bibitem[{Hinsen(2018)Hinsen, Konrad}]{hinsen2018verifiability}
\textbf{\color{eLifeMediumGrey} Hinsen K}.
\newblock Verifiability in computer-aided research: the role of digital
  scientific notations at the human-computer interface.
\newblock PeerJ Computer Science.  2018; 4:e158.

\bibitem[{Hsieh et~al.(2018)Hsieh, Terry and Vaickus, Max H and Remick, Daniel
  G}]{hsieh2018enhancing}
\textbf{\color{eLifeMediumGrey} Hsieh T}, Vaickus MH, Remick DG.
\newblock Enhancing scientific foundations to ensure reproducibility: a new
  paradigm.
\newblock The American journal of pathology.  2018; 188(1):6--10.

\bibitem[{Hunter(2017)Hunter, Philip}]{hunter2017the}
\textbf{\color{eLifeMediumGrey} Hunter P}.
\newblock The reproducibility “crisis”.
\newblock EMBO reports.  2017; 18(9):1493--1496.
\newblock
  \urlprefix\url{https://www.embopress.org/doi/abs/10.15252/embr.201744876},
  \href{10.15252/embr.201744876}{\doiprefix
  \detokenize{10.15252/embr.201744876}}.

\bibitem[{Hussain et~al.(2013)Hussain, Waqar and Moens, Nathalie and Veraitch,
  Farlan S and Hernandez, Diana and Mason, Chris and Lye, Gary
  J}]{hussain2013reproducible}
\textbf{\color{eLifeMediumGrey} Hussain W}, Moens N, Veraitch FS, Hernandez D,
  Mason C, Lye GJ.
\newblock Reproducible Culture and Differentiation of Mouse Embryonic Stem
  Cells Using an Automated Microwell Platform.
\newblock Biochemical engineering journal.  2013; 77(100):246--257.

\bibitem[{Hutson(2018)Hutson, Matthew}]{hutson2018artificial}
\textbf{\color{eLifeMediumGrey} Hutson M}.
\newblock Artificial intelligence faces reproducibility crisis.
\newblock Science.  2018; 359(6377):725--726.
\newblock \urlprefix\url{https://science.sciencemag.org/content/359/6377/725},
  \href{10.1126/science.359.6377.725}{\doiprefix
  \detokenize{10.1126/science.359.6377.725}}.

\bibitem[{Jamieson et~al.(2019)Jamieson, Kathleen Hall and McNutt, Marcia and
  Kiermer, Veronique and Sever, Richard}]{jamieson2019Signaling}
\textbf{\color{eLifeMediumGrey} Jamieson KH}, McNutt M, Kiermer V, Sever R.
\newblock Signaling the trustworthiness of science.
\newblock Proceedings of the National Academy of Sciences.  2019;
  116(39):19231--19236.
\newblock \urlprefix\url{https://www.pnas.org/content/116/39/19231},
  \href{10.1073/pnas.1913039116}{\doiprefix
  \detokenize{10.1073/pnas.1913039116}}.

\bibitem[{{P}roject {J}upyter et~al.(2018){P}roject {J}upyter and Matthias
  Bussonnier and Jessica Forde and J. Freeman and Brian E. Granger and T. Head
  and Chris Holdgraf and K. Kelley and Gladys Nalvarte and Andrew Osheroff and
  M. Pacer and Yuvi Panda and Fernando P{\'e}rez and Benjamin Ragan-Kelley and
  Carol Willing}]{jupyter2018binder}
\textbf{\color{eLifeMediumGrey} {P}roject {J}upyter}, Bussonnier M, Forde J,
  Freeman J, Granger BE, Head T, Holdgraf C, Kelley K, Nalvarte G, Osheroff A,
  Pacer M, Panda Y, P{\'e}rez F, Ragan-Kelley B, Willing C.
\newblock {B}inder 2.0 - {R}eproducible, interactive, sharable environments for
  science at scale.
\newblock In: \emph{{P}roceedings of the 17th {P}ython in {S}cience
  {C}onference}; 2018. p. 113 -- 120.
\newblock \href{10.25080/Majora-4af1f417-011}{\doiprefix
  \detokenize{10.25080/Majora-4af1f417-011}}.

\bibitem[{Kelly(2019)Kelly, Clint D}]{kelly2019rate}
\textbf{\color{eLifeMediumGrey} Kelly CD}.
\newblock Rate and success of study replication in ecology and evolution.
\newblock PeerJ.  2019 sep 10; 7:e7654.
\newblock \href{10.7717/peerj.7654}{\doiprefix
  \detokenize{10.7717/peerj.7654}}.

\bibitem[{Kluyver et~al.(2016)Thomas Kluyver and Benjamin Ragan-Kelley and
  Fernando P{\'e}rez and Brian Granger and Matthias Bussonnier and Jonathan
  Frederic and Kyle Kelley and Jessica Hamrick and Jason Grout and Sylvain
  Corlay and Paul Ivanov and Dami{\'a}n Avila and Safia Abdalla and Carol
  Willing and the Jupyter development team}]{kluyver2016jupyter}
\textbf{\color{eLifeMediumGrey} Kluyver T}, Ragan-Kelley B, P{\'e}rez F,
  Granger B, Bussonnier M, Frederic J, Kelley K, Hamrick J, Grout J, Corlay S,
  Ivanov P, Avila D, Abdalla S, Willing C, the Jupyter~development team.
\newblock Jupyter Notebooks-a publishing format for reproducible computational
  workflows.
\newblock In: \emph{ELPUB}; 2016. p. 87--90.

\bibitem[{Kroeger et~al.(2018)Kroeger, Cynthia M and Garza, Cutberto and Lynch,
  Christopher J and Myers, Esther and Rowe, Sylvia and Schneeman, Barbara O and
  Sharma, Arya M and Allison, David B}]{kroeger2018scientific}
\textbf{\color{eLifeMediumGrey} Kroeger CM}, Garza C, Lynch CJ, Myers E, Rowe
  S, Schneeman BO, Sharma AM, Allison DB.
\newblock Scientific rigor and credibility in the nutrition research landscape.
\newblock The American journal of clinical nutrition.  2018 March;
  107(3):484—494.
\newblock \urlprefix\url{https://europepmc.org/articles/PMC6248649},
  \href{10.1093/ajcn/nqx067}{\doiprefix \detokenize{10.1093/ajcn/nqx067}}.

\bibitem[{Lannelongue et~al.(2021{\natexlab{a}})Lannelongue, Loïc and Grealey,
  Jason and Bateman, Alex and Inouye, Michael}]{lannelongue2021ten}
\textbf{\color{eLifeMediumGrey} Lannelongue L}, Grealey J, Bateman A, Inouye M.
\newblock Ten simple rules to make your computing more environmentally
  sustainable.
\newblock PLOS Computational Biology.  2021 sep 20; 17(9):e1009324.

\bibitem[{Lannelongue et~al.(2021{\natexlab{b}})Lannelongue, Loïc and Grealey,
  Jason and Inouye, Michael}]{lannelongue2021green}
\textbf{\color{eLifeMediumGrey} Lannelongue L}, Grealey J, Inouye M.
\newblock Green Algorithms: Quantifying the Carbon Footprint of Computation.
\newblock Advanced Science.  2021; n/a(n/a):2100707.
\newblock
  \urlprefix\url{https://onlinelibrary.wiley.com/doi/abs/10.1002/advs.202100707},
  \href{https://doi.org/10.1002/advs.202100707}{\doiprefix
  \detokenize{https://doi.org/10.1002/advs.202100707}}.

\bibitem[{Ledermann and Gartner(2021)Ledermann, Florian and Gartner,
  Georg}]{ledermann2021towards}
\textbf{\color{eLifeMediumGrey} Ledermann F}, Gartner G.
\newblock Towards {Conducting} {Reproducible} {Distributed} {Experiments} in
  the {Geosciences}.
\newblock AGILE: GIScience Series.  2021 jun 4; 2:1--7.
\newblock \href{10.5194/agile-giss-2-33-2021}{\doiprefix
  \detokenize{10.5194/agile-giss-2-33-2021}}.

\bibitem[{Mayr et~al.(2020)Mayr, Fabian and M{\" o}ller, Gabriele and Garscha,
  Ulrike and Fischer, Jana and Casta{\~ n}o, Patricia Rodr{\' i}guez and
  Inderbinen, Silvia G and Temml, Veronika and Waltenberger, Birgit and
  Schwaiger, Stefan and Hartmann, Rolf W and Gege, Christian and Martens,
  Stefan and Odermatt, Alex and Pandey, Amit V. and Werz, Oliver and Adamski,
  Jerzy and Stuppner, Hermann and Schuster, Daniela}]{mayr2020finding}
\textbf{\color{eLifeMediumGrey} Mayr F}, M{\" o}ller G, Garscha U, Fischer J,
  Casta{\~ n}o PR, Inderbinen SG, Temml V, Waltenberger B, Schwaiger S,
  Hartmann RW, Gege C, Martens S, Odermatt A, Pandey AV, Werz O, Adamski J,
  Stuppner H, Schuster D.
\newblock Finding {New} {Molecular} {Targets} of {Familiar} {Natural}
  {Products} {Using} {In} {Silico} {Target} {Prediction}.
\newblock International Journal of Molecular Sciences.  2020 sep 26; 21(19).

\bibitem[{Meng(2020)Meng, Xiao-Li}]{meng2020reproducibility}
\textbf{\color{eLifeMediumGrey} Meng XL}.
\newblock Reproducibility, {Replicability}, and {Reliability}.
\newblock Harvard Data Science Review.  2020 oct 29; 2(4).
\newblock Https://hdsr.mitpress.mit.edu/pub/hn51kn68.

\bibitem[{Meyerowitz-Katz et~al.(2021)Meyerowitz-Katz, Gideon and Besan{\c
  c}on, Lonni and Flahault, Antoine and Wimmer, Raphael}]{meyerowitz2021impact}
\textbf{\color{eLifeMediumGrey} Meyerowitz-Katz G}, Besan{\c c}on L, Flahault
  A, Wimmer R.
\newblock Impact of mobility reduction on {COVID}-19 mortality: absence of
  evidence might be due to methodological issues.
\newblock Scientific Reports.  2021 dec 7; 11(1).

\bibitem[{N{\" a}pflin et~al.(2019)N{\" a}pflin, Kathrin and O'Connor, Emily A
  and Becks, Lutz and Bensch, Staffan and Ellis, Vincenzo A and Hafer-Hahmann,
  Nina and Harding, Karin C and Lind{\' e}n, Sara K and Olsen, Morten T and
  Roved, Jacob and Sackton, Timothy B and Shultz, Allison J and
  Venkatakrishnan, Vignesh and Videvall, Elin and Westerdahl, Helena and
  Winternitz, Jamie C and Edwards, Scott V}]{napflin2019genomics}
\textbf{\color{eLifeMediumGrey} N{\" a}pflin K}, O'Connor EA, Becks L, Bensch
  S, Ellis VA, Hafer-Hahmann N, Harding KC, Lind{\' e}n SK, Olsen MT, Roved J,
  Sackton TB, Shultz AJ, Venkatakrishnan V, Videvall E, Westerdahl H,
  Winternitz JC, Edwards SV, Genomics of host-pathogen interactions: challenges
  and opportunities across ecological and spatiotemporal scales; 2019.

\bibitem[{Nielsen et~al.(2017)Nielsen, Finn {\r A}rup and Mietchen, Daniel and
  Willighagen, Egon}]{nielsen2017scholia}
\textbf{\color{eLifeMediumGrey} Nielsen F{\r A}}, Mietchen D, Willighagen E.
\newblock Scholia, {Scientometrics} and {Wikidata}.
\newblock In: \emph{The {Semantic} {Web}: ESWC 2017 {Satellite} {Events}};
  2017. p. 237--259.
\newblock \href{10.1007/978-3-319-70407-4_36}{\doiprefix
  \detokenize{10.1007/978-3-319-70407-4_36}}.

\bibitem[{N{\" u}st et~al.(2020)N{\" u}st, Daniel and Sochat, Vanessa V. and
  Marwick, Ben and Eglen, Stephen J. and Head, Tim and Hirst, Tony and Evans,
  Benjamin D}]{nust2020ten}
\textbf{\color{eLifeMediumGrey} N{\" u}st D}, Sochat VV, Marwick B, Eglen SJ,
  Head T, Hirst T, Evans BD.
\newblock Ten simple rules for writing {Dockerfiles} for reproducible data
  science.
\newblock PLOS Computational Biology.  2020 nov 10; 16(11):e1008316.
\newblock \href{10.1371/JOURNAL.PCBI.1008316}{\doiprefix
  \detokenize{10.1371/JOURNAL.PCBI.1008316}}.

\bibitem[{Peng(2015)Peng, Roger}]{peng2015thereproducibility}
\textbf{\color{eLifeMediumGrey} Peng R}.
\newblock The reproducibility crisis in science: A statistical counterattack.
\newblock Significance.  2015; 12(3):30--32.

\bibitem[{Pimentel et~al.(2019)Pimentel, Jo\~{a}o Felipe and Murta, Leonardo
  and Braganholo, Vanessa and Freire, Juliana}]{pimentel2019a}
\textbf{\color{eLifeMediumGrey} Pimentel JaF}, Murta L, Braganholo V, Freire J.
\newblock A Large-scale Study About Quality and Reproducibility of Jupyter
  Notebooks.
\newblock In: \emph{Proceedings of the 16th International Conference on Mining
  Software Repositories} MSR '19, Piscataway, NJ, USA: IEEE Press; 2019. p.
  507--517.
\newblock \href{10.1109/MSR.2019.00077}{\doiprefix
  \detokenize{10.1109/MSR.2019.00077}}.

\bibitem[{Pimentel et~al.(2021)Jo{\~{a}}o Felipe Pimentel and Leonardo Murta
  and Vanessa Braganholo and Juliana Freire}]{pimentel2021understanding}
\textbf{\color{eLifeMediumGrey} Pimentel JF}, Murta L, Braganholo V, Freire J.
\newblock Understanding and improving the quality and reproducibility of
  Jupyter notebooks.
\newblock Empir Softw Eng.  2021; 26(4):65.
\newblock \urlprefix\url{https://doi.org/10.1007/s10664-021-09961-9},
  \href{10.1007/s10664-021-09961-9}{\doiprefix
  \detokenize{10.1007/s10664-021-09961-9}}.

\bibitem[{Plesser(2017)Plesser, Hans E}]{plesser2017reproducibility}
\textbf{\color{eLifeMediumGrey} Plesser HE}.
\newblock Reproducibility vs. {Replicability}: A {Brief} {History} of a
  {Confused} {Terminology}.
\newblock Frontiers in Neuroinformatics.  2017 jan 1; 11:76.

\bibitem[{{Project Jupyter}(2021)}]{nbdime}
\textbf{\color{eLifeMediumGrey} {Project Jupyter}}, nbdime: Jupyter Notebook
  Diff and Merge tools; 2021.
\newblock Accessed 18 May 2021.
\newblock \url{https://github.com/jupyter/nbdime}.

\bibitem[{Randles et~al.(2017)Randles, Bernadette M and Pasquetto, Irene V and
  Golshan, Milena S and Borgman, Christine L}]{randles2017using}
\textbf{\color{eLifeMediumGrey} Randles BM}, Pasquetto IV, Golshan MS, Borgman
  CL.
\newblock Using the Jupyter notebook as a tool for open science: An empirical
  study.
\newblock In: \emph{2017 ACM/IEEE Joint Conference on Digital Libraries (JCDL)}
  IEEE; 2017. p. 1--2.

\bibitem[{Roberts(2001)Roberts, Richard J.}]{roberts2001pubmed}
\textbf{\color{eLifeMediumGrey} Roberts RJ}.
\newblock PubMed Central: The GenBank of the published literature.
\newblock Proceedings of the National Academy of Sciences.  2001;
  98(2):381--382.
\newblock \urlprefix\url{https://www.pnas.org/content/98/2/381},
  \href{10.1073/pnas.98.2.381}{\doiprefix \detokenize{10.1073/pnas.98.2.381}}.

\bibitem[{Rule et~al.(2019)Rule, A and Birmingham, A and Zuniga, C and
  Altintas, I and Huang, SC and Knight, R and Moshiri, N and Nguyen, MH and
  Rosenthal, SB and P{\'e}rez, F and others}]{rule2019ten}
\textbf{\color{eLifeMediumGrey} Rule A}, Birmingham A, Zuniga C, Altintas I,
  Huang S, Knight R, Moshiri N, Nguyen M, Rosenthal S, P{\'e}rez F, et~al.
\newblock Ten simple rules for writing and sharing computational analyses in
  Jupyter Notebooks.
\newblock Plos Computational Biology.  2019; 15(7):e1007007--e1007007.

\bibitem[{Rule et~al.(2018)Rule, Adam and Tabard, Aur{\'e}lien and Hollan,
  James D.}]{rule2018exploration}
\textbf{\color{eLifeMediumGrey} Rule A}, Tabard A, Hollan JD.
\newblock Exploration and Explanation in Computational Notebooks.
\newblock In: \emph{Proceedings of the 2018 CHI Conference on Human Factors in
  Computing Systems} CHI '18, New York, NY, USA: ACM; 2018. p. 32:1--32:12.
\newblock \href{10.1145/3173574.3173606}{\doiprefix
  \detokenize{10.1145/3173574.3173606}}.

\bibitem[{Russell et~al.(2018)Russell, Pamela H and Johnson, Rachel L and
  Ananthan, Shreyas and Harnke, Benjamin and Carlson, Nichole
  E}]{russell2018large}
\textbf{\color{eLifeMediumGrey} Russell PH}, Johnson RL, Ananthan S, Harnke B,
  Carlson NE.
\newblock A large-scale analysis of bioinformatics code on {GitHub}.
\newblock PLOS ONE.  2018; 13(10):e0205898.

\bibitem[{Rutz et~al.(2022)Rutz, Adriano and Sorokina, Maria and Galgonek,
  Jakub and Mietchen, Daniel and Willighagen, Egon and Gaudry, Arnaud and
  Graham, James G and Stephan, Ralf and Page, Roderic and Vondr{\' a}{\v s}ek,
  Ji{\v r}{\' i} and Steinbeck, Christoph and Pauli, Guido F and Wolfender,
  Jean-Luc and Bisson, Jonathan and Allard, Pierre-Marie}]{rutz2022LOTUS}
\textbf{\color{eLifeMediumGrey} Rutz A}, Sorokina M, Galgonek J, Mietchen D,
  Willighagen E, Gaudry A, Graham JG, Stephan R, Page R, Vondr{\' a}{\v s}ek J,
  Steinbeck C, Pauli GF, Wolfender JL, Bisson J, Allard PM.
\newblock The {LOTUS} initiative for open knowledge management in natural
  products research.
\newblock eLife.  2022 may 26; 11.

\bibitem[{Samuel and K{\"{o}}nig{-}Ries(2018)Sheeba Samuel and Birgitta
  K{\"{o}}nig{-}Ries}]{samuel2018provbook}
\textbf{\color{eLifeMediumGrey} Samuel S}, K{\"{o}}nig{-}Ries B.
\newblock ProvBook: Provenance-based Semantic Enrichment of Interactive
  Notebooks for Reproducibility.
\newblock In: van Erp M, Atre M, L{\'{o}}pez V, Srinivas K, Fortuna C, editors.
  \emph{Proceedings of the {ISWC} 2018 Posters {\&} Demonstrations, Industry
  and Blue Sky Ideas Tracks co-located with 17th International Semantic Web
  Conference {(ISWC} 2018), Monterey, USA, October 8th - to - 12th, 2018}, vol.
  2180 of {CEUR} Workshop Proceedings CEUR-WS.org; 2018.
  \urlprefix\url{http://ceur-ws.org/Vol-2180/paper-57.pdf}.

\bibitem[{Samuel and K{\"o}nig-Ries(2021)Samuel, Sheeba and K{\"o}nig-Ries,
  Birgitta}]{samuel2021reproducemegit}
\textbf{\color{eLifeMediumGrey} Samuel S}, K{\"o}nig-Ries B.
\newblock ReproduceMeGit: A Visualization Tool for Analyzing Reproducibility of
  Jupyter Notebooks.
\newblock In: Glavic B, Braganholo V, Koop D, editors. \emph{Provenance and
  Annotation of Data and Processes} Cham: Springer International Publishing;
  2021. p. 201--206.

\bibitem[{Samuel and König-Ries(2021)Samuel, Sheeba and König-Ries,
  Birgitta}]{samuel2021understanding}
\textbf{\color{eLifeMediumGrey} Samuel S}, König-Ries B.
\newblock Understanding experiments and research practices for reproducibility:
  an exploratory study.
\newblock PeerJ.  2021 Apr; 9:e11140.
\newblock \urlprefix\url{https://doi.org/10.7717/peerj.11140},
  \href{10.7717/peerj.11140}{\doiprefix \detokenize{10.7717/peerj.11140}}.

\bibitem[{Samuel et~al.(2020)Samuel, Sheeba and Shadaydeh, Maha and B{\"
  o}cker, Sebastian and Br{\" u}gmann, Bernd and Bucher, Solveig Franziska and
  Deckert, Volker and Denzler, Joachim and Dittrich, Peter and von Eggeling,
  Ferdinand and G{\" u}llmar, Daniel and Guntinas-Lichius, Orlando and K{\"
  o}nig-Ries, Birgitta and L{\" o}ffler, Frank and Maicher, Lutz and Marz,
  Manja and Migliavacca, Mirco and Reichenbach, J{\" u}rgen R. and Reichstein,
  Markus and R{\" o}mermann, Christine and Wittig, Andrea}]{samuel2020virtual}
\textbf{\color{eLifeMediumGrey} Samuel S}, Shadaydeh M, B{\" o}cker S, Br{\"
  u}gmann B, Bucher SF, Deckert V, Denzler J, Dittrich P, von Eggeling F, G{\"
  u}llmar D, Guntinas-Lichius O, K{\" o}nig-Ries B, L{\" o}ffler F, Maicher L,
  Marz M, Migliavacca M, Reichenbach JR, Reichstein M, R{\" o}mermann C, Wittig
  A.
\newblock A virtual ``{Werkstatt}'' for digitization in the sciences.
\newblock Research Ideas and Outcomes.  2020 may 11; 6.

\bibitem[{Sandve et~al.(2013)Sandve, Geir Kjetil AND Nekrutenko, Anton AND
  Taylor, James AND Hovig, Eivind}]{sandve2013ten}
\textbf{\color{eLifeMediumGrey} Sandve GK}, Nekrutenko A, Taylor J, Hovig E.
\newblock Ten Simple Rules for Reproducible Computational Research.
\newblock PLOS Computational Biology.  2013 10; 9(10):1--4.
\newblock \urlprefix\url{https://doi.org/10.1371/journal.pcbi.1003285},
  \href{10.1371/journal.pcbi.1003285}{\doiprefix
  \detokenize{10.1371/journal.pcbi.1003285}}.

\bibitem[{Sayers(2010)Sayers, Eric}]{sayers2010ageneral}
\textbf{\color{eLifeMediumGrey} Sayers E}.
\newblock A General Introduction to the E-utilities.
\newblock Entrez Programming Utilities Help [Internet] Bethesda (MD): National
  Center for Biotechnology Information (US).  2010; .

\bibitem[{Schr{\"{o}}der et~al.(2019)Max Schr{\"{o}}der and Frank Kr{\"{u}}ger
  and Sascha Spors}]{schroder2019reproducible}
\textbf{\color{eLifeMediumGrey} Schr{\"{o}}der M}, Kr{\"{u}}ger F, Spors S.
\newblock Reproducible Research is more than Publishing Research Artefacts: {A}
  Systematic Analysis of Jupyter Notebooks from Research Articles.
\newblock CoRR.  2019; abs/1905.00092.
\newblock \urlprefix\url{http://arxiv.org/abs/1905.00092}.

\bibitem[{Schwartz et~al.(2020)Schwartz, Roy and Dodge, Jesse and Smith, Noah
  A. and Etzioni, Oren}]{schwartz2020green}
\textbf{\color{eLifeMediumGrey} Schwartz R}, Dodge J, Smith NA, Etzioni O.
\newblock Green {AI}.
\newblock Communications of the ACM.  2020 nov 17; 63(12):54--63.

\bibitem[{Shepperd et~al.(2018)Shepperd, Martin and Ajienka, Nemitari and
  Counsell, Steve}]{shepperd2018role}
\textbf{\color{eLifeMediumGrey} Shepperd M}, Ajienka N, Counsell S.
\newblock The role and value of replication in empirical software engineering
  results.
\newblock Information and Software Technology.  2018 7; 99:120--132.

\bibitem[{Siebert et~al.(2015)Siebert, Sabina and Machesky, Laura M and Insall,
  Robert H}]{siebert2015point}
\textbf{\color{eLifeMediumGrey} Siebert S}, Machesky LM, Insall RH.
\newblock Point of view: Overflow in science and its implications for trust.
\newblock Elife.  2015; 4:e10825.

\bibitem[{Simmons et~al.(2011)Simmons, J. P. and Nelson, L. D. and Simonsohn,
  U.}]{simmons2011false}
\textbf{\color{eLifeMediumGrey} Simmons JP}, Nelson LD, Simonsohn U.
\newblock False-{Positive} {Psychology}: Undisclosed {Flexibility} in {Data}
  {Collection} and {Analysis} {Allows} {Presenting} {Anything} as
  {Significant}.
\newblock Psychological Science.  2011 oct 17; 22(11):1359--1366.
\newblock \href{10.1177/0956797611417632}{\doiprefix
  \detokenize{10.1177/0956797611417632}}.

\bibitem[{Taddeo et~al.(2021)Taddeo, Mariarosaria and Tsamados, Andreas and
  Cowls, Josh and Floridi, Luciano}]{taddeo2021artificial}
\textbf{\color{eLifeMediumGrey} Taddeo M}, Tsamados A, Cowls J, Floridi L.
\newblock Artificial intelligence and the climate emergency: Opportunities,
  challenges, and recommendations.
\newblock One Earth.  2021; 4(6):776--779.

\bibitem[{{The Economist}(2013)}]{theeconomist2013trouble}
\textbf{\color{eLifeMediumGrey} {The Economist}}, Trouble at the lab; 2013.
\newblock
  \urlprefix\url{https://www.economist.com/briefing/2013/10/18/trouble-at-the-lab}.

\bibitem[{Trisovic et~al.(2022)Trisovic, Ana and Lau, Matthew K and Pasquier,
  Thomas and Crosas, Merc{\` e}}]{trisovic2022large}
\textbf{\color{eLifeMediumGrey} Trisovic A}, Lau MK, Pasquier T, Crosas M.
\newblock A large-scale study on research code quality and execution.
\newblock Scientific Data.  2022 feb 21; 9(1):60.

\bibitem[{Vargo and Gilbert(2020)Vargo, Alexander H S and Gilbert, Anna
  C}]{vargo2020rank}
\textbf{\color{eLifeMediumGrey} Vargo AHS}, Gilbert AC.
\newblock A rank-based marker selection method for high throughput {scRNA}-seq
  data.
\newblock BMC Bioinformatics.  2020 oct 23; 21(1):477.

\bibitem[{Waagmeester et~al.(2020)Waagmeester, Andra and Stupp, Gregory and
  Burgstaller-Muehlbacher, Sebastian and Good, Benjamin M. and Griffith,
  Malachi and Griffith, Obi and Hanspers, Kristina and Hermjakob, Henning and
  Hudson, Toby and Hybiske, Kevin and Keating, Sarah M and Manske, Magnus and
  Mayers, Michael and Mietchen, Daniel and Mitraka, Elvira and Pico, Alexander
  R. and Putman, Timothy Elliott and Riutta, Anders and Rosinach, N{\' u}ria
  Queralt and Schriml, Lynn and Shafee, Thomas and Slenter, Denise and Stephan,
  Ralf and Thornton, Katherine and Tsueng, Ginger and Tu, Roger and Ul-Hasan,
  Sabah and Willighagen, Egon and Wu, Chunlei and Su, Andrew
  I.}]{waagmeester2020wikidata}
\textbf{\color{eLifeMediumGrey} Waagmeester A}, Stupp G,
  Burgstaller-Muehlbacher S, Good BM, Griffith M, Griffith O, Hanspers K,
  Hermjakob H, Hudson T, Hybiske K, Keating SM, Manske M, Mayers M, Mietchen D,
  Mitraka E, Pico AR, Putman TE, Riutta A, Rosinach NQ, Schriml L, et~al.
\newblock Wikidata as a knowledge graph for the life sciences.
\newblock eLife.  2020 mar 17; 9.

\bibitem[{Wang et~al.(2020{\natexlab{a}})Wang, Jiawei and Kuo, Tzu-yang and Li,
  Li and Zeller, Andreas}]{wang2020restoring}
\textbf{\color{eLifeMediumGrey} Wang J}, Kuo Ty, Li L, Zeller A.
\newblock Restoring Reproducibility of Jupyter Notebooks.
\newblock In: \emph{2020 IEEE/ACM 42nd International Conference on Software
  Engineering: Companion Proceedings (ICSE-Companion)}; 2020. p. 288--289.

\bibitem[{Wang et~al.(2020{\natexlab{b}})Wang, Jiawei and Li, Li and Zeller,
  Andreas}]{wang2020better}
\textbf{\color{eLifeMediumGrey} Wang J}, Li L, Zeller A.
\newblock Better code, better sharing: on the need of analyzing jupyter
  notebooks.
\newblock In: \emph{Proceedings of the ACM/IEEE 42nd International Conference
  on Software Engineering: New Ideas and Emerging Results}; 2020. p. 53--56.

\bibitem[{Willcox(2021)Willcox, Aaron}]{Willcox2021ReSearchOps}
\textbf{\color{eLifeMediumGrey} Willcox A}, ReSearchOps: a principled framework
  and guide to computational reproducibility.
\newblock Open Science Framework; 2021.

\bibitem[{Willis et~al.(2020)Willis, Alistair and Charlton, Patricia and Hirst,
  Tony}]{willis2020developing}
\textbf{\color{eLifeMediumGrey} Willis A}, Charlton P, Hirst T.
\newblock Developing students' written communication skills with Jupyter
  notebooks.
\newblock In: \emph{Proceedings of the 51st ACM Technical Symposium on Computer
  Science Education}; 2020. p. 1089--1095.

\bibitem[{Wofford et~al.(2019)Wofford, Morgan F and Boscoe, Bernadette M and
  Borgman, Christine L and Pasquetto, Irene V and Golshan, Milena
  S}]{wofford2019jupyter}
\textbf{\color{eLifeMediumGrey} Wofford MF}, Boscoe BM, Borgman CL, Pasquetto
  IV, Golshan MS.
\newblock Jupyter notebooks as discovery mechanisms for open science: Citation
  practices in the astronomy community.
\newblock Computing in Science \& Engineering.  2019; 22(1):5--15.

\bibitem[{Woodbridge(2017)Mark Woodbridge}]{woodbridge2017jupyter}
\textbf{\color{eLifeMediumGrey} Woodbridge M}, Jupyter Notebooks and
  reproducible data science; 2017.
\newblock
  \urlprefix\url{https://markwoodbridge.com/2017/03/05/jupyter-reproducible-science.html}.

\bibitem[{Zhu et~al.(2021)Zhu, Chenguang and Saha, Ripon K. and Prasad, Mukul
  R. and Khurshid, Sarfraz}]{zhu2021restoring}
\textbf{\color{eLifeMediumGrey} Zhu C}, Saha RK, Prasad MR, Khurshid S.
\newblock Restoring the Executability of Jupyter Notebooks by Automatic Upgrade
  of Deprecated APIs.
\newblock In: \emph{2021 36th IEEE/ACM International Conference on Automated
  Software Engineering (ASE)}; 2021. p. 240--252.
\newblock \href{10.1109/ASE51524.2021.9678889}{\doiprefix
  \detokenize{10.1109/ASE51524.2021.9678889}}.

\end{thebibliography}

\end{document}